\documentstyle[sprocl]{article}
\input epsf.tex
\def\DESepsf(#1 width #2){\epsfxsize=#2 \epsfbox{#1}}
\def\be{\begin{equation}}
\def\ee{\end{equation}}
\def\bea{\begin{eqnarray}}
\def\eea{\end{eqnarray}}

\begin{document}
\title{CP Violation
\footnote{ Lecture given at the CCAST workshop on "CP 
Violation and Various Frontiers in Tau and Other Systems", CCAST, Beijing,
China, 11-14, August, 1997.}
}  
\author{ Xiao-Gang He }
\address{School of Physics, University of Melbourne\\
Parkville, Vic. 3052, Australia}
\author{  }
\address{ }
\maketitle\abstracts{
In this lecture I review the present status of 
$CP$ violation in the Standard Model and some of its 
extensions and discuss ways to distinguish different models. 
}
{\small{
{\bf Contents}\\

1.  { Introduction}





2.  { CP Violation In The Standard Model}



3.  { Test The Standard Model in $B$ Decays}






4.  { Models For $CP$ Violation}



5.  { The KM Unitarity Triangle And New Physics}

6.  { Direct $CP$ Violation In Neutral Kaon System}

7.  { The Electric Dipole Moment}




8.  { Partial Rate Asymmetry}

9.  { Test Of $CP$ Violation Involving Polarisation Measurement}




10.{ Baryon Number Asymmetry}

11.{ Conclusion}
}} 

\newpage
\section{Introduction}

Symmetries play  very important role in physics. 
They often simplify the analyses of  complex systems. 
These symmetries may be continuos or discrete.
For each symmetry there is a corresponding conservation law~\cite{noether}.
In the real physical world, some of the symmetries are exact and some are 
broken. The studies of symmetries conserved as well as broken ones 
are all important. These studies  have provided 
many insights for the understanding of the fundamental principles of 
the universe.

Different interactions in nature have different symmetry properties. 
Experiments have not found any violation of 
energy-momentum conservation and angular momentum 
conservation in all 
known interactions (gravity, strong, and electroweak interactions).
These are the consequences of exact 
continuos space-time symmetry (Translational and Lorentz invariance).
One can also define discrete space-time symmetries, such as:
spatial inversion symmetry $P$ (the parity symmetry), 
the time reversal symmetry
$T$, and the charge conjugation symmetry $C$ (the particle and anti-particle 
symmetry). The last one is
related to discrete space-time symmetry in the sense that an anti-particle 
can be viewed as a particle moving "backwards" 
in time due to~\cite{stu-feyn}
St\" uckelberg and Feynman.
 
For many years, $C$, $P$ and $T$, were thought to be separately conserved
in all interactions. 
This believe was proven to be wrong in the mid 50's.
In 1956, Lee and Yang first proposed that parity $P$
might not be conserved in weak interactions~\cite{lee-yang}. 
Shortly thereafter, $P$ 
violation was experimentally confirmed in nucleon 
$\beta$ decays~\cite{wu} and in $\pi$ and $\mu$ decays~\cite{parity}.
This opened a new chapter in elementary particle 
physics and led to a major advance in the understanding of weak interactions.
In 1964 Christenson, Cronin, Fitch and Turlay made another advance. They 
discovered
that the combined $CP$ symmetry was also violated in weak decays of neutral
kaons~\cite{cp-exp}. 
They found that about $0.2\%$ of the long lived neutral kaon $K_L$, 
thought to be 
 a particle with $CP = -1$ would decay into two $\pi$ final state, 
a state with $CP = +1$. Up to now this is the only laboratory system in which 
$CP$ violation has been observed. 
There have been 
many theoretical attempts trying 
to understand the origin of $CP$ violation, but so far there is no
satisfactory explanation~\cite{cp-rev}. 
In this lecture, I will review some of the recent 
developements in the study of CP violation.

Let me begin with some basics about the discrete space-time symmetries,
$C$, $P$ and $T$.

\subsection{ Parity Symmetry}

The parity operation is a spatial inversion through the origin, a mirror 
reflection. 
Mathematically,
the effect of parity operation on the wave function $\psi(\vec x)$ of a 
state $|N, \vec p, \vec s>$, (Here $N$ refers to internal quantum numbers, 
such as: electric charge, baryon number and etc., $\vec p$ and $\vec s$ are the
momentum and spin, respectively), can be expressed as

\begin{eqnarray}
P \psi(\vec x) \rightarrow \psi (-\vec x)\;.
\label{eparity}
\end{eqnarray}
It has the effect of reversing momenta but leaving spins and other internal quantum
numbers unchanged:

\begin{eqnarray}
P |N,\;\vec p,\;\vec s> = \eta_P | N,\; -\vec p,\;\vec s>\;,
\end{eqnarray}
where $\eta_P$ is a phase factor which is identified with the intrinsic parity.
The intrinsic parity of a particle can be determined by first 
assigning intrinsic parity to proton, neutron etc., and then study their 
strong interactions with the particle in question.

In quantum mechanics, the parity symmetry (invariance) 
of the interactions, i.e. the
property that the interaction potential $V(\vec x)$ is unchanged by 
the parity operation 
\begin{eqnarray}
V(\vec x) = V(-\vec x)\;,
\end{eqnarray}
implies $\psi(-\vec x)\psi^*(-\vec x)$ is equal to $\psi(\vec x)\psi^*(\vec x)$,
and therefore 
the
consequence that

{\it{The probability of the transition $i\rightarrow f$ is the same as the
probability for $Pi \rightarrow Pf$, where $Pi$ and $Pf$ are the parity 
transformed states of $i$ and $f$.}}

\begin{figure}[htb]
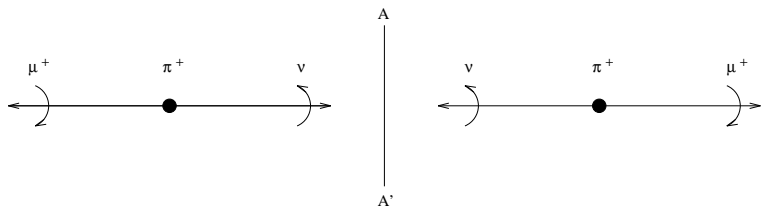

\centerline{ \DESepsf(hefig1.epsf width 10 cm) }
\caption {Mirror processes.}
\label{parity}
\end{figure}

A great advance in the understanding of weak interactions came in 1956
when it was discovered that weak interactions are not invariant under parity
transformation~\cite{lee-yang,wu,parity}.
The basic idea can be illustrated by one of the classic experiments
which established parity violation in weak interactions --
observations of the decay $\pi^+ \rightarrow \mu^+  \nu$~\cite{parity}. 
Suppose the initial
pion ($\pi^+$) is at rest. It has zero spatial momentum and zero angular 
momentum, 
the latter because the $\pi^+$ being a pseudoscalar 
has no intrinsic spin. The muon ($\mu^+$) and
the neutrino ($\nu$) each have an intrinsic spin of 1/2. 
The final state must also have
zero total spatial momentum and zero total angular momentum.
A possible configuration is that shown in the left half of
Fig.\ref{parity}, in which the muon and the neutrino are travelling ``back to back'',
both the muon and the neutrino having ``left-handed'' angular momentum about the
direction
of motion. 
Another possible state is illustrated in the right 
half of Fig.\ref{parity}, which is obtained by reflecting
the left state in the ``mirror''
represented by the line AA' in the diagram. 
In this second possible final state
the neutrino and the muon are both right-handed. While this second state is
theoretically possible, in that it is consistent with the laws of conservation
of linear and angular momentum, it is not observed in nature. As the state on
the left is observed and the one on the right is not, it clearly indicates
that the weak interaction is not invariant under the parity transformation.

This lack of $P$ invariance is most succinctly expressed by saying that weak
interactions involve only left-handed neutrinos. $P$ invariance would require
equal coupling to left-handed and right-handed neutrinos, so one sees that there
is a maximum violation of parity symmetry in weak interactions. The
essential left-handedness of weak interactions was an important clue which
led to the present understanding of weak interactions. 

\subsection{ Time Reversal Symmetry}

Classically time reversal operation $T$ corresponds to the operation:
 $t \rightarrow -t$.
This has the effect of reversing momenta, spins and interchanging 
the initial and final states. 

The non-invariance of macroscopic dynamics under the reversal of the direction
of time is well known and
is often illustrated by running a movie backwards. Here another
example is given, a damped pendulum.
The transformation $T$ applied to the equation of motion for
the damped pendulum
\begin{eqnarray}
m{d^2x\over d^2t} + r {dx\over dt} + k x =0\;
\end{eqnarray}
gives the $T$ transformed equation
\begin{eqnarray}
m{d^2x\over d^2t} - r {dx\over dt} + k x =0\;
\end{eqnarray}
which describes growing rather than decaying oscillations. Clearly the first
order derivative in the equation of motion  is the reason why it 
is not $T$ invariant.

In quantum mechanics, the situation is more complicated. The 
Schr\"odinger equation
\begin{eqnarray}
i\hbar {d\psi\over dt} = [ -{\hbar^2\over 2m} \nabla^2 + V(t)]\psi\;,
\label{sch}
\end{eqnarray}
is an equation with first derivative in time. However it 
can still be made $T$ invariant. This seems to be in contradiction with the 
observation for the damped pendulum. 
This puzzle was solved by Wigner in 1932~\cite{wigner}. 
The $T$ operation is not a simple 
sign change in time in quantum mechanics. It is a combined transformation:

{\it{Change $t$ to $-t$ and take the complex conjugate.}}

Thus the $T$ transformed version of eq.(\ref{sch}) is
\begin{eqnarray}
i\hbar {d\psi^*\over dt} = [-{\hbar^2\over 2 m} \nabla^2 + V^*(t)]
\psi^*\;,\label{sch1}
\end{eqnarray}
but as quantum observables are expectation values involving only $\psi^*\psi$,
as long as the interaction $V$ is real (i.e. $V^* =V$), eqs.(\ref{sch}) 
and (\ref{sch1}) 
describe the same physics. In other words time reversal invariance in quantum
mechanics imposes reality conditions on the interaction.
To break $T$ symmetry in a quantum system one needs  to introduce
complex valued interactions somehow.

\subsection{Charge Conjugation Symmetry}

Charge conjugation is an 
operation which takes a particle into its anti-particle.
Application of the charge conjugation 
changes the signs of all additive quantum numbers,
but leaves particle spins and momenta unchanged, that is

\begin{eqnarray}
C|N,\; \vec p,\;\vec s> = \eta_C|-N,\;\vec p,\;\vec s>\;.
\end{eqnarray}
Here $\eta_C$ is a phase factor. It can be easily seen that a particle or a 
particle system is a charge 
conjugation eigenstate only if its additive quantum 
numbers
are all zero. An example is the $\pi^0$ which satisfies:

\begin{eqnarray}
C|\pi^0> = + |\pi^0>\;.
\end{eqnarray}
Such a property is known as self-conjugate.

Charge conjugation symmetry also plays an important role in particle physics.
Let me return to the pion decay reaction
\begin{eqnarray}
\pi^+ \rightarrow \mu^+ \nu\;,
\end{eqnarray}
and replace each particle by its anti-particle, so that the reaction becomes
\begin{eqnarray}
\pi^-\rightarrow \mu^- \bar \nu\;.
\end{eqnarray}

\begin{figure}[htb]
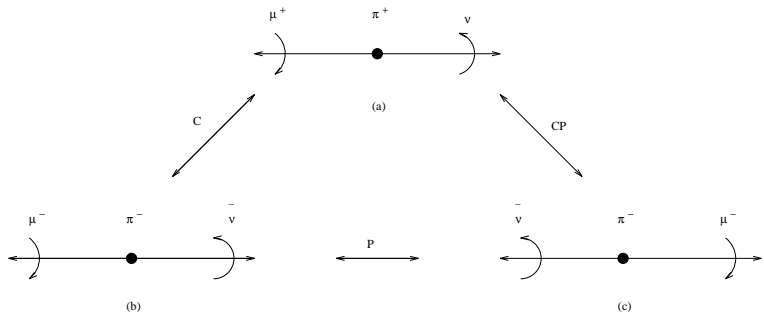

\centerline{ \DESepsf(hefig3.epsf width 10 cm) }
\caption {C, P and CP transformed processes.}
\label{charge}
\end{figure}

The situation is depicted in Fig.\ref{charge}.
The initial reaction is shown in Fig.\ref{charge}(a), and the effect of
replacing particles by their anti-particles is illustrated in Fig.\ref{charge}
(b).
However the reaction in Fig.\ref{charge}(b) was not observed, 
but reaction illustrated in Fig.\ref{charge}(c) was observed,
which may be obtained from Fig.\ref{charge}(b) by a mirror reflection.
This implies that 
weak interactions are not $P$ invariant or $C$ invariant, but are invariant
under the combined $CP$ transformation.

$CP$ was still considered to be an exact symmetry.
At least that was believed to be the case until 1964 when it was discovered
that the weak interactions responsible for the decays of neutral kaons 
into pions 
were not exactly $CP$ invariant either~\cite{cp-exp}. To
understand this result it is necessary to look at the peculiar properties of
the neutral kaon system. 

Under a $P$ transformation, because kaons and pions are pseudoscalar, one has
\begin{eqnarray}
\pi\rightarrow -\pi\;,\;\;
K\rightarrow -K\;,
\end{eqnarray}

Under a $C$ transformation one can choose a convention such that 
\begin{eqnarray}
\pi^+\rightarrow \pi^-\;,\;\;
\pi^0\rightarrow \pi^0\;,\;\;
K^0\rightarrow \bar K^0\;.
\end{eqnarray}

While the neutral pion is its own anti-particle, the neutral kaon is not, its
two varieties $K^0$ and $\bar K^0$ being distinguished by their strangeness
quantum numbers, $S = +1 $ for $K^0$, and $S = -1$ for $\bar K^0$.

The neutral kaons decay into two and three pions by weak interactions.
If $CP$ is conserved one would expect that the interactions responsible for 
these decays will connect 
states with the same $CP$ eigenvalues.
There are two neutral kaon $CP$ eigenstates which can
be constructed from
$K^0$ and $\bar K^0$,
\begin{eqnarray}
K^0_1 &=& {1\over \sqrt{2}}(K^0-\bar K^0)\;,\;\;\;\;CP\;\;\mbox{even}\;;
\nonumber\\
K^0_2&=& {1\over \sqrt{2}}(K^0+\bar K^0)\;,\;\;\;\;CP\;\;\mbox{odd}\;.
\end{eqnarray}

The pion systems from neutral kaon decays are determined from experiments to be 
in states with no relative angular momenta between pions (S-wave states). The
two pion systems $(\pi^+\pi^-\;,\pi^0\pi^0)$ and the three pion systems
$(\pi^+\pi^-\pi^0\;,\pi^0\pi^0\pi^0)$ are in $CP$ even and odd states, respectively.
Thus the expected decays are
\begin{eqnarray}
&&K^0_1\rightarrow \pi^+\pi^-\;,\; \pi^0\pi^0\;,\nonumber\\
&&K^0_2 \rightarrow \pi^+\pi^-\pi^0\;,\; \pi^0\pi^0\pi^0\;.
\end{eqnarray}

The masses of the pion and kaon are about 140 MeV and 490 MeV
respectively, so that there is much less energy available for the $3\pi$ decays
than there is for the $2\pi$ decays, and kinematic arguments suggest that the
$K^0_2$ decay will be much slower than the $K^0_1$ decay. This is the case, the
observed lifetimes being about $10^{-7}$s and $10^{-10}$s, respectively. A
consequence of this is that, simply by waiting long enough a neutral kaon beam
will become a pure $K^0_2$ beam, expected to decay to three pions. But in 1964
it was observed that about a few in a thousand long-lived kaons decayed into
two pions~\cite{cp-exp}. 
This suggests that the long-lived kaon $K_L$ and the short-live 
kaon $K_S$ are admixture of $K_1^0$ and
$K_2^0$ (or $K^0$ and  $\bar K^0$ mixing). This is usually expressed as
\begin{eqnarray}
K_L = {K_2^0+\epsilon_1 K_1^0\over \sqrt{1+|\epsilon_1|^2}}\;,\;\;
K_S = {K_1^0+\epsilon_2 K_2^0\over \sqrt{1+|\epsilon_2|^2}}\;.
\end{eqnarray}
Here the most general parameterization have been used 
allowing $\epsilon_1$ to
be different from $\epsilon_2$. They are of the order $10^{-3}$.
It is clear that weak 
interactions violate $CP$ symmetry, but do so weakly, unlike the
maximal violation of $P$ symmetry. 

\subsection{The $CPT$ Theorem}

As have been discussed in previous sections that 
discrete symmetries $C$ and  $P$ by itself is not 
conserved. The same applies to the product symmetry $CP$.
What about the triple product symmetry $CPT$? So far there is no experimental
evidence which shows the violation of this symmetry. 
In fact there is more foundamental reason why $CPT$ symmetry should be 
conserved. 
In the 1950s, it was shown~\cite{cpt} 
that $CPT$ is always conserved in the framework
of a local quantum field theory with Lorentz invariance, Hermiticity 
and the usual 
spin-statistics (Bose-Einstein statistics for bosons, 
and Fermi-Dirac statistics 
for fermions). This is the so
called $CPT$ theorem.

There are many implications of the $CPT$ theorem. For example, 
the masses, and
life-times are all equal for particles and their corresponding anti-particles. 
These properties provide
practical ways to test the $CPT$ theorem. Let me now discuss 
the implications of the 
$CPT$ theorem for 
$K^0-\bar K^0$ system~\cite{t-cpt}.

The weak interaction connecting the two neutral kaons $K^0$ and $\bar K^0$
 in the system can be
parameterized in quantum mechanics by an effective Hamiltonian $H$. 
In general it contains
two Hermitian $2\times2$ matrices $M$ and $\Gamma$,
\begin{eqnarray}
H = M - i{\Gamma\over 2} =\left ( \begin{array}{ll}
M_{11}-i\Gamma_{11}/2&M_{12}-i\Gamma_{12}/2\\
M_{12}^*-i\Gamma_{12}^*/2&M_{22}-i\Gamma_{22}/2
\end{array}
\right )\;,
\end{eqnarray}
where $\Gamma$ is related to the
life-times, and $M$ is related to the masses of the particles.
In the basis of $(K^0\;,\;\;\bar K^0)$, the diagonal entries
$M_{11,22}$ and $\Gamma_{11,22}$ are the masses and life-times of 
$K^0$ and $\bar K^0$, respectively. The off diagonal ones mix $K^0$ and
$\bar K^0$. If $CPT$ symmetry is exact, $M_{11} = M_{22}$ and $\Gamma_{11}
=\Gamma_{22}$. 

Allowing $M_{11}$ and $\Gamma_{11}$ to be different from $M_{22}$ and 
$\Gamma_{22}$ explicitly violates $CPT$ symmetry. 
Different experiments can be 
performed to test $CPT$ symmetry. So far all experimental results are 
consistent with the assumption that $CPT$ is an exact symmetry. 
The best limit on 
$CPT$ symmetry is from the mass difference between the masses of $K^0$ and
$\bar K^0$, one has~\cite{pdg} 
$|m_{K^0} - m_{\bar K^0}|/m_{K^0} < 9\times 10^{-19}$.
However, at present only partial aspects of the $CPT$ theorem have been tested.
One should keep an open mind about the validity of the $CPT$ theorem. In the
lack of evidence for $CPT$ violation,
$CPT$ symmetry will be assumed to be an exact symmetry in later discussions.

With $CPT$ symmetry for $K^0-\bar K^0$ system, 
one obtains the mass and life-time eigenvalues
for $K_S$ and $K_L$ 
\begin{eqnarray}
(m-i{\Gamma\over 2})_S = M_{11} - i{\Gamma_{11}\over 2} - E\;,\nonumber\\
(m-i{\Gamma\over 2})_L = M_{11} - i{\Gamma_{11}\over 2} + E\;,\nonumber\\
E = \sqrt{(M_{12}-i\Gamma_{12}/2)(M_{12}^*-i\Gamma^*_{12}/2)}\;.
\end{eqnarray}
In this case, $\epsilon_1$ is equal to $\epsilon_2$ which will be denoted by
$\epsilon$. One obtains:
\begin{eqnarray}
\epsilon &=& {\sqrt{M_{12}-i\Gamma_{12}/2} - \sqrt{M_{12}^*-i\Gamma_{12}^*/2}
\over \sqrt{M_{12}-i\Gamma_{12}/2} + \sqrt{M_{12}^*-i\Gamma_{12}^*/2}}\nonumber\\
&\approx&
{iIm(M_{12}) + Im(\Gamma_{12}/2)\over \Delta m_{L-S} +i\Delta_{S-L}/2}\;.
\end{eqnarray}
Here $\Delta m_{L-S}
= m_L-m_S$ and $\Delta \Gamma_{S-L} = \Gamma_S - \Gamma_L$. 
$\epsilon$ is experimentally measured~\cite{pdg} 
to be $2.27\times 10^{-3}exp(i\phi_\epsilon)$
with $\phi_\epsilon = 46.3^o$.

Using the facts:
$\Delta m_{L-S} \approx \Delta \Gamma_{S-L}/2$, and $Im(\Gamma_{12})$ is 
much smaller than $Im(M_{12})$ from theoretical estimate, one finally obtains
\begin{eqnarray}
\epsilon \approx {Im(M_{12})\over \sqrt{2} \Delta m_{L-S}} e^{i\phi_\epsilon}\;,
\end{eqnarray} 

To understand $CP$ violation, one must understand
how $Im(M_{12})$ is generated and what is the origin of it. 
The key to the question is to have complex interactions.
However, there are many possible 
ways to introduce complex interactions. 

Many possible explanations~\cite{super,spon,weinberg,l-r,km} for CP violation 
in neutral kaon system 
have been put forward since its surprising discovery in 1964. 
One of the early popular model is the superweak model. 
This model assumes that there
is a new complex interaction which changes the strange number by two units with
a strength approximately $10^{-10}$ weaker than the standard weak interaction. 
It causes the mixing between $K^0$ and $\bar K^0$ with the
right order of magnitude. If one assumes that 
the coupling of the new interaction 
is the same order of magnitude as the standard weak interaction, the energy
scale of the new physics would be at the order of $10000$ TeV. 
Other mechanisms for $CP$ violation include 
phases in the left-handed charged current, 
phases in the 
right-handed charged current, phases in the vacuum expectation values and etc..
I will discuss some of these models in the following sections.

\section{CP Violation In The Standard Model}

\subsection{The Kobayashi-Moskawa Model}

In the SM the strong and electroweak interactions are described by
$SU(3)_C\times SU(2)_L \times U(1)_Y$ gauge interactions~\cite{w-sm,s-sm}. The 
$SU(3)_C$ gauge interaction describes the strong interaction 
whose gauge bosons are the eight gluons~\cite{s-sm}. 
The $SU(2)_L\times
U(1)_Y$ gauge interactions describe the electroweak interactions~\cite{w-sm}. 
The
corresponding gauge bosons are $W^\pm$, $Z$ and $\gamma$. 
The matter fields are
the left-handed leptons 
$L_L$, the 
right-handed charged leptons $E_R$, the left-handed quarks $Q_L$, the
right-handed up quarks $U_R$, and down quarks $D_R$. 
Their transformation properties 
under the SM gauge group are:

\begin{eqnarray}
L_L: (1, 2, -1)\;;&& E_R: (1, 1, -2)\;;\nonumber\\
 Q_L: (3,2,1/3);&& U_R: (3,1, 4/3)\;; 
D_R: (3,1,-2/3)
\;.
\end{eqnarray}
Each of such a set is called a generation. Three generations have been 
experimentally observed. Experimental data from LEP~\cite{lep}  as well as 
from nucleosynthesis~\cite{nucl-syn} show that there are only 
three light nuetrino generations. 

The SM gauge group $SU(3)_C\times SU(2)_L \times U(1)_Y$ is broken down to
$SU(3)_C\times U(1)_{em}$ at about $100$ GeV. 
Before symmetry breaking, the gauge
bosons and matter fields are all massless. After symmetry breaking, 
three of the gauge bosons $W^\pm$, $Z$ and the charged
leptons and quarks become massive. The mechanism for electroweak symmetry
breaking is not well understood and is  an 
outstanding problem of particle
physics~\cite{breaking}. 
In the SM, the symmetry breaking is
due to the vacuum expectation value (VEV) 
$<H> = v$ of a Higgs doublet $H: (1,2,-1)$. This is the Higgs
 mechanism~\cite{higgs}. This
model predicts the existence of a neutral scalar particle with mass less than
a $TeV$ or so if the Higgs sector is weakly coupled~\cite{higgs-mass}.
The other three degrees of freedom of $H$ are "eaten" by the $W^\pm$ and $Z$
after symmetry breaking.
The matter fields obtain their masses from their Yukawa 
couplings to $H$.
The couplings  are given by

\begin{eqnarray}
L_Y = \bar Q_{Li} \lambda^U_{ij}H U_{Rj} + \bar Q_{Li} \lambda^D_{ij} \tilde
H D_{Rj} + \bar L_{Li} \lambda^E_{ij} \tilde H E_{Rj} + H.C.
\end{eqnarray}
where $\tilde H = i\sigma_2 H^*$, and 
the subindices $i$ and $j$ are the generation indices.
In this model neutrinos are still massless after symmetry breaking. Because 
of this fact 
the matrix $\lambda^E$ can be rotated into a diagonal form without
loss of generality. However, the diagonalization of the matrices
$M^{U,D} = v \lambda^{U,D}$ will not be trivial.   
This is related to $CP$ violation in the SM which was first realised by
Kobayashi and 
Moskawa in 1973 (the KM mechanism)~\cite{km}.
This mechanism is called the SM for $CP$ violaiton. 
In this model $CP$ violation arises from the
complex phases in the charged current of weak interactions due to miss
match of the weak and mass eigenstates of quarks. 

In the weak interaction eigenstate basis, the charged current is given by

\begin{eqnarray}
L_W = {g\over \sqrt{2}} \bar U_L \gamma_\mu D_L W^\mu + H.C.
\end{eqnarray}
where $U_L = (u,c,t,...)_L$ and $D_L = (d,s,b,...)_L$.

In the quark mass
eigenstate basis the charged current interaction becomes,

\begin{eqnarray}
L_W = {g\over \sqrt{2}} \bar U^m_L V_{KM}\gamma^\mu D^m_L W_\mu + H.C.
\end{eqnarray}
where $U_L^m = V_L^UU_L$, and $D^m_L = V_L^D D_L$, and $V_{KM} = V_L^U V_L^{D\dagger}$ is called the Kobayashi-Moskawa matrix. Here $V_L^{U,D}$ are 
unitary matrices which diagonalize the mass matrices,

\begin{eqnarray}
V^U_L M^U V^U_R &=& diag(m_u,\;m_c,\;m_t,\;...)\;,\nonumber\\
V^D_L M^D V^D_R &=& diag(m_d,\;m_s,\;m_b,\;...)\;.
\end{eqnarray} 
The KM matrix $V_{KM}$ 
is an $N\times N$ unitary 
matrix which contains $N^2$ parameters for $N$ generations.
 Among the $N^2$ parameters
$2N-1$ parameters can be absorbed into the redefinition of quark
phases and therefore are not physical ones. 
The remaining matrix is
described by $N(N-1)/2$ rotation angles, and $(N-1)(N-2)/2$ phases.
Non-zero values for the phases are the sources for CP violation in the SM.
It is easily seen that in order to have CP violation, there should 
exist at least three
generations. 
The original parameterization of $V_{KM}$ with three generations 
due to Kobayashi and
Moskawa is given by~\cite{km}

\begin{eqnarray}
V_{KM} &=&
\left ( \begin{array}{lll}
V_{ud}&V_{us}&V_{ub}\\
V_{cd}&V_{cs}&V_{cb}\\
V_{td}&V_{ts}&V_{tb}
\end{array}
\right )\nonumber\\
&=& \left ( \begin{array}{lll}
c_1&-s_1c_3& -s_1s_3\\
s_1c_2&c_1c_2c_3-s_2s_3e^{i\delta}&c_1c_2s_3+s_2c_3e^{i\delta}\\
s_1s_2&c_1s_2c_3+c_2s_3e^{i\delta}&c_1s_2s_3-c_2c_3e^{i\delta}
\end{array}
\right )\;, 
\end{eqnarray}
where $s_i = sin\theta_i$ and $c_i = cos\theta_i$ with $\theta_i$ being the 
rotation angles. A non-zero value for $\delta$ violates $CP$.
In many cases it is convenient to use the Wolfenstein 
parameterization~\cite{wolf} which is
give by 

\begin{eqnarray}
V_{KM} \approx \left ( \begin{array}{lll}
1-\lambda^2/2& \lambda& A\lambda^3(\rho - i \eta)\\
-\lambda& 1-\lambda^2/2 & A\lambda^2\\
A\lambda^3(1-\rho - i\eta)&-A\lambda^2& 1
\end{array}
\right )\;.
\end{eqnarray}
When discussing $CP$ violation, it is necessary to keep higher order terms in 
$\lambda$, one should add $-A^2\lambda^5(\rho+i\eta)$ and 
$-A\lambda^4(\rho+i\eta)$ to $V_{cd}$ and $V_{ts}$, respectively.
$CP$ violation in this parameterization is characterised by a non-zero value
for $\eta$. 

The magnitudes for the KM matrix elements $|V_{ij}|$ are constrained by several 
experiments. They are summarised in the following:

\begin{eqnarray}
\begin{array}{ll}
|V_{ud}| = 0.9836\pm0.0010
& \mbox{From comparison of nuclear $\beta$}\\
& \mbox{decays to muon decay~\cite{vud};}\\
&\\
|V_{us}| = \lambda =0.2205\pm 0.0018&\mbox{From  $K_{e3}$ and hyperon 
decays~\cite{vus};}\\
&\\
|V_{cb}| = 0.0393\pm 0.0028
&\mbox{From $B\rightarrow D^* l\bar \nu_l$ and inclusive $B$ 
decays~\cite{vcb};}\\
 \rightarrow A = 0.808\pm 0.058&\\
&\\
|V_{ub}/V_{cb}| = 0.08\pm 0.016
& \mbox{$b$ to $u$ transition~\cite{vub}.}
\\
\rightarrow (\rho^2+\eta^2)^{1/2} 
= 0.363\pm 0.073&
\end{array}
\end{eqnarray}

Without considering $CP$ violating experimental data, it is not possible to 
separately determine $\eta$ and $\rho$. 

\subsection{$CP$ violation in $K^0-\bar K^0$ mixing}

\begin{figure}[htb]
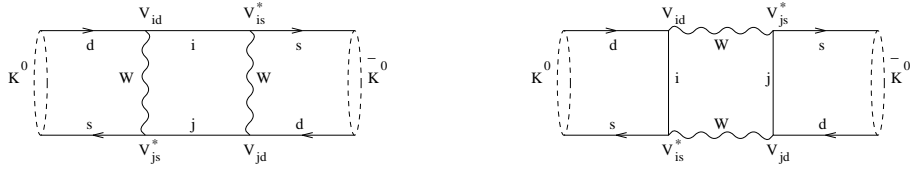

\centerline{ \DESepsf(hefig4.epsf width 12 cm) }
\caption {"box" diagrams for $K^0 - \bar K^0$ mixing in the Standard Model.}
\label{mixing}
\end{figure}

In the SM, the mixing of $K^0$ and $\bar K^0$ occurs at one loop level 
as shown in Fig.\ref{mixing}~\cite{box}.
Evaluating these Feynman diagrams, one obtains the $\Delta S = 2$ 
effective Hamiltonian~\cite{box},
\begin{eqnarray}
H_{eff} &=& -{2\over 3} {G_F^2m_W^2\over \pi^2} \sum_{i,j} (V_{id}V_{is}^*)
(V_{jd}V_{js}^*) B(\alpha_i, \alpha_j) \bar s \gamma_\mu L d 
\bar s \gamma^\mu L d\;,\nonumber\\
B(x,y)&=& (1+{xy\over 4}) ( {1\over (1-x)(1-y)} +
{1\over x-y}[ {x^2ln x\over (1-x)^2} - {y^2lny\over (1-y)^2}])\nonumber\\
&-& 2xy({1\over (1-x)(1-y)} + {1\over x-y} [ {xlnx\over (1-x)^2} -
{ylny\over (1-y)^2}])\;,
\end{eqnarray}
and 
$\alpha_i =m_i^2/ m_W^2$.
 
The transition matrix element $M_{12}$ is given by
\begin{eqnarray}
M_{12} &=& <\bar K^0|H_{eff}| K^0>\nonumber\\
&=& 
-{2\over 3} {G_F^2m_W^2\over \pi^2} \sum_{i,j} (V_{id}V_{is}^*)
(V_{jd}V_{js}^*) B(\alpha_i, \alpha_j)C\;,
\end{eqnarray}
where $C = <\bar K^0|\bar s\gamma_\mu L d \bar s\gamma^\mu L d|K^0>$. In
the vacuum saturation approximation,

\begin{eqnarray}
C 
= <\bar K^0|\bar s \gamma_\mu L d|0><0|\bar s \gamma^\mu L d |K^0>\nonumber=
 - {1\over 8} f_K^2 m_K\;,
\end{eqnarray}
where $f_K = 160$ MeV is the kaon decay constant.
To take into account of other contributions, one introduces a parameter $B_K$,
some times called the bag factor, such that
\begin{eqnarray}
C = -{1\over 8} f_K^2 m_K B_K\;.
\end{eqnarray}
There are many estimates for 
this parameter. In the  numerical calculation, 
$B_{K} = 0.75 \pm 0.15$ will be used~\cite{rosner}.

With QCD corrections, the matrix element $M_{12}$ is given by

\begin{eqnarray}
M_{12} &=& {f_K^2m_K G_F^2 m_W^2\over 12\pi^2} B_K
[\eta_1 \tilde B_1 (V_{cd}V_{cs}^*)^2 +\eta_2 \tilde B_2 (V_{td}V_{ts}^*)^2
\nonumber\\
&+&2\eta_3 \tilde B_3 (V_{cd}V_{cs}^* V_{td}V_{ts}^*)]\;,\nonumber\\
\tilde B_1 &=& B(\alpha_c,\alpha_c) - B(\alpha_u,\alpha_c) -B(\alpha_c,\alpha_u)
+B(\alpha_u,\alpha_u)\;,\nonumber\\
\tilde B_2&=& B(\alpha_t,\alpha_t) - B(\alpha_u,\alpha_t) -B(\alpha_t,\alpha_u)
+B(\alpha_u,\alpha_u)\;,\nonumber\\
\tilde B_3 &=& B(\alpha_u,\alpha_u) 
-B(\alpha_c,\alpha_u) - B(\alpha_t,\alpha_u)
+B(\alpha_t, \alpha_c)\;,
\end{eqnarray}
where the QCD correction factors $\eta_i$ have been evaluated up 
to next-to-leading
order and are given by: $\eta_1 = 1.38$, $\eta_2 = 0.57$, and 
$\eta_3 = 0.47$~\cite{qcd}.

The parameter $\epsilon$ is given by
\begin{eqnarray}
&&|\epsilon| = 4.39 A^2 B_K \eta [\eta_3 \tilde B_3 - \eta_1 \tilde B_1
+ \eta_2 A^2 \lambda^4(1-\rho) \tilde B_2]\;.
\label{eeta}
\end{eqnarray}

Experimental data from $B^0-\bar B^0$ mixing provides additional constraint
on the parameter. Evaluating similar diagrams as in Fig.\ref{mixing}
 for $B^0-\bar B^0$ 
mixing, one has
\begin{eqnarray}
\Delta m = {f_B^2 m_B G_F^2m_W^2\over 6 \pi^2}B_B \eta_B\tilde B_2 
|V_{td}V_{tb}^*|^2\;,
\label{bb}
\end{eqnarray}
where $B_B$ is the bag factor for $B^0-\bar B^0$ mixing, $\eta_B$ is the QCD 
correction factor which is equal to $0.55$~\cite{qcd}. 

Inserting the values: $m_t(m_W) = 165\pm6$ GeV~\cite{top}, 
$\Delta m/\Gamma_B = 0.73
\pm 0.05$~\cite{pdg} 
and 
$f_B\sqrt{B_B} = 200\pm 40$ MeV~\cite{rosner} 
in eqs. (\ref{eeta}) and (\ref{bb}), 
one obtains,

\begin{eqnarray}
&&\eta (1-\rho + 0.44) = 0.51 \pm 0.18\;,\nonumber\\
&&(1-\rho)^2-\eta^2 = 1.02\pm 0.44\;.
\end{eqnarray}

Combining information from
$|V_{ub}/V_{cb}|$ and the above two equations, 
one finally obtains the allowed region for $\rho$ and $\eta$,

\begin{eqnarray}
-0.24 <\rho < 0.24\;,\;\;0.18<\eta <0.42\;.
\end{eqnarray}
The SM is consistent
with experimental data.


Of course fitting $\epsilon$ alone 
is not enough to establish the SM for CP violation. 
More experiments 
should be performed to test the SM. 
$CP$ violation experiments to be carried out at
$B$ factories will provide excellent opportunities 
to test the SM model which will be discussed
in the following section.

\section{Test The SM In B Decays}

\begin{figure}[htb]
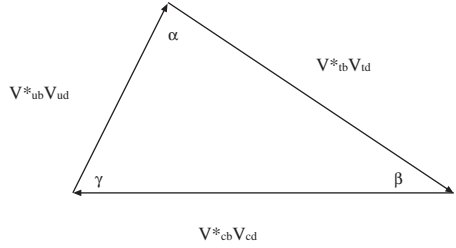

\centerline{ \DESepsf(CKMTriangle.epsf width 6 cm) }
\caption {The KM unitarity triangle.}
\label{triangle}
\end{figure}

An unique feature of the SM for $CP$ 
violation is that the 
KM matrix is a $3\times3$ unitary matrix. Due to the unitarity property, 
when summed over 
the row or column of matrix elements $V_{ij}$ times complex conjugate matrix 
elements $V_{ik}^*$, the following equations hold,
\begin{eqnarray}
\sum_i V_{ij}V^*_{ik} = \delta_{jk}\;,\;\;\;\sum_i V_{ji}V_{ki}^* = \delta_{jk}\;.
\label{tri}
\end{eqnarray}
These equations define six triangles when $j \neq k$. 
For example, for $j = d$ and $k = b$ a triangle shown in Fig.\ref{triangle}
  with three angles
$\alpha = \mbox{Arg}(-V_{td}V^*_{tb}/V_{ub}^*V_{ud})$,
$\beta = \mbox{Arg}(-V_{cd}V^*_{cb}/V_{tb}^*V_{td})$, and
$\gamma =  \mbox{Arg}(-V_{ud}V^*_{ub}/V_{cb}^*V_{cd})$ is defined. 
The angles from the six triangles mentioned earlier 
completely determine the KM matrix. Only
four angles are independent~\cite{k-l}. Among the three angles $\alpha$, 
$\beta$ and $\gamma$, two of them are independent because 
$\alpha+\beta +\gamma = 180^o$.
The other
two independent angles can be chosen to be: 
$\sigma = \mbox{Arg}(-V_{cs}^*V_{cb}/V_{ts}^*V_{tb})$ and $\sigma' 
= \mbox{Arg}(-V_{ud}^*V_{us}/V_{cd}^*V_{cs})$.
Present experimental data constrain the angles $\sigma$ and $\sigma'$ to
be very small compared with the angles $\alpha$, $\beta$ and $\gamma$.

Among the six triangles defined by eq.(\ref{tri}), the one in 
Fig.\ref{triangle} will be
experimentally studied in the near future. 
If $CP$ is conserved in the KM sector the triangle shrinks to a line. To test
the KM mechanism for $CP$ violation it is sufficient to measure the three
angles $\alpha$, $\beta$ and $\gamma$ and 
to see if they add up to $180^o$. Many methods have been proposed to 
determine these angles~\cite{methods}.
Alternatively 
one can also
test the KM mechanism by measuring: 1) Two angles and one
ratio of two sides of the triangle, for example, $|V_{ub}^*V_{ud}/V_{cb}^*V_{cd}|$;
2) One angle and two ratios for different 
two sides; and 3) Three ratios for different 
two sides. 

\subsection{The effective Hamiltonian for $B$ decays}

\begin{figure}[htb]
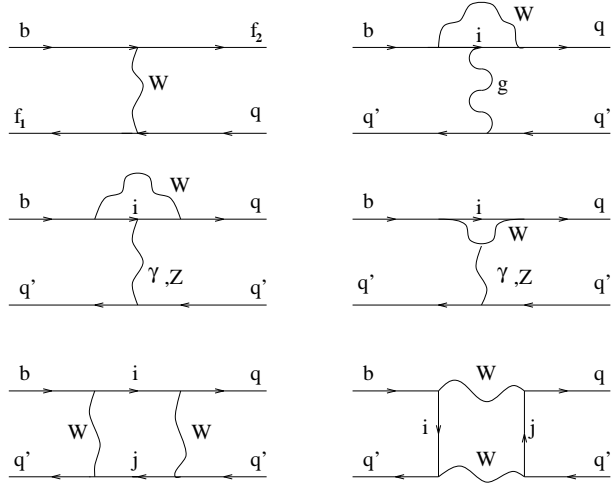

\centerline{ \DESepsf(fig-sm-decay.epsf width 8 cm) }
\caption {The Feynman diagrams for $B$ decays up to one loop in electroweak 
interactions.}
\label{decay}
\end{figure}

In this section the effective Hamiltonian responsible for $B$ decays will be
given.
Both tree and loop contributions to $B$ decays are important. The Feynman 
diagrams for these decays up to one loop in electroweak interactions are shown 
in Fig.\ref{decay}.

 The effective Hamiltonian obtained from these diagrams with QCD corrections 
can be written as~\cite{h-b,d-h1} 
\begin{eqnarray}
 H_{\Delta B =1}(q) &&= {4 G_{F} \over \sqrt{2}} [
V_{ub}V_{cq}^*(c_1 O_{1}^{uc}(q) + c_2 O_2^{uc}(q))
\nonumber\\
&&+V_{cb}V_{uq}^*(c_1 O^{cu}_1(q) + c_2 O^{cu}_2(q))
+V_{ub}V^{*}_{uq} (c_1 O^{u}_1(q)\nonumber\\
&& +c_2 O^{u}_2(q)) 
+ V_{cb}V^{*}_{cq} (c_1 O^{c}_1(q) +c_2 O^{c}_2(q))
\nonumber\\
  && - 
\sum _{j=u,c,t}V_{jb}V^{*}_{jq} \sum_{i=3}^{10} c_{i}^j O_{i}(q)] + H.C. ,
\end{eqnarray} 
where $O_{i}$'s are defined as 
\begin{eqnarray}
 &&O^{f_1 f_2}_{1}(q) = \bar q_{\alpha} \gamma_{\mu} L f_{1\beta} \bar f_{2\beta} \gamma^{\mu} L b_{\alpha} ,
  \ \  O^{f_1f_2}_{2}(q) = \bar q \gamma_{\mu} L f_1 \bar f_2 \gamma^{\mu} L b ,           \nonumber \\ 
 &&O^{f}_{1}(q) = \bar q_{\alpha} \gamma_{\mu} L f_{\beta} \bar f_{\beta} \gamma^{\mu} L b_{\alpha} ,
  \ \  O^{f}_{2}(q) = \bar q \gamma_{\mu} L f \bar f \gamma^{\mu} L b ,           \\ 
 &&O_{3(5)}(q) = \bar q \gamma_{\mu} L b \bar q^{\prime} \gamma^{\mu} L(R) q^{\prime} ,   
  \ \ O_{4(6)}(q) = \bar q_{\alpha} \gamma_{\mu} L b_{\beta} \bar q^{\prime}_{\beta} 
       \gamma^{\mu} L(R) q^{\prime}_{\alpha}  ,                         \nonumber \\ 
 &&O_{7(9)}(q) = {3 \over 2} \bar q \gamma_{\mu} L b e_{q^{\prime}} \bar q^{\prime} 
       \gamma^{\mu} R(L) q^{\prime} ,        
  \ \ O_{8(10)}(q) ={3 \over 2} \bar q_{\alpha} \gamma_{\mu} L b_{\beta}  e_{q^{\prime}} 
       \bar q^{\prime}_{\beta} \gamma^{\mu} R(L) q^{\prime}_{\alpha} ,\nonumber      
\end{eqnarray} 
where $f$ can be $u$ or $c$ quark, $q$ can be $d$ or $s$ quark, 
and $q^{\prime}$ is summed over $u$, $d$, $s$, and $c$ quarks.  $\alpha$ and $\beta$ are 
the color indices.  
$T^{a}$ is the SU(3) generator with the normalisation $Tr(T^{a} T^{b}) = \delta^{ab}/2$.  
$G^{\mu \nu}_{a}$ and $F_{\mu \nu}$ are the gluon and photon field strengths, 
respectively. $c_i$ 
are the Wilson Coefficients (WC).  
The next-to-leading order QCD corrected WC's 
with 
$\alpha_{s}(m_{Z}) =0.118$, $\alpha_{em}(m_{Z}) =1/128$, $m_{t}=176$ GeV and 
$\mu \approx m_{b}=5$ GeV, are given by~\cite{d-h1} 
\begin{eqnarray}
 c_1 &=& -0.3125,  \ \  c_2 = 1.1502,  \ \  c_3^t = 0.0174,  \ \  c_4^t = -0.0373,  \nonumber \\ 
 c_5^t &=& 0.0104,  \ \  c_6^t = -0.0459,  \ \  c_7^t = -1.050 \times 10^{-5},  \nonumber \\ 
 c_8^t &=& 3.839 \times 10^{-4},  \ \  c_9^t = -0.0101,  \ \  c_{10}^t = 1.959 \times 10^{-3}\;,\nonumber\\
c_{3,5}^{u,c}&=& - c_{4,6}^{u,c}/N_c = P_s^{u,c}/N_c\;,
c_{7,9}^{u,c} = P_{em}^{u,c}\;,\;\; c_{8,10}^{u,c} = 0\;,
\end{eqnarray} 
where $N_c$ is the number of color, $P_s^i = (\alpha_s/8\pi) c_2
[10/9 + G(m_i,\mu,q^2)]$, and $P_{em}^i = \alpha_{em}/9\pi) ( N_c c_1 + c_2) 
[10/9 + G(m_i, \mu, q^2)]$. 
The function $G(m,\mu, q^2) = 4 \int^1_0 x(1-x) ln[(m^2-x(1-x)q^2)/\mu^2]dx$.

%

One expects that the hadronic matrix elements arising from quark operator $O_{1-10}$ to be the same order of magnitudes. The relevant strengths of the 
contribution from each term in $H_{eff}$ are predominantly determined by their
corresponding KM factors and the WC's. This will provide guidance to identify 
dominant contribution for a decay process.

\subsection{The determination of the angle $\alpha$}
At asymmetric  $B$ factories it is possible to measure the time variation of rate
asymmetries of $B$ and $\bar B$. This provides an excellent opportunity to
determine some of the angles~\cite{b-factory,stone}.
As an example 
let me first consider the standard method to measure $\alpha$ in 
$B\rightarrow \pi\pi$~\cite{sanda,gl}. 
The time-dependent rate for initially pure $B^0$ or $\bar B^0$ to decay into a final 
CP eigenstate, for example $\pi^{+} \pi^{-}$ at time $t$ is~\cite{sanda}
\begin{eqnarray} 
 \Gamma(t) &=& |A|^2 e^{-\Gamma t} [{1+|\xi |^2 \over 2} 
    + {1-|\xi |^2 \over 2} \cos{(\Delta M t)} - Im \xi \sin{(\Delta M t)}] , \nonumber \\ 
\bar  \Gamma(t) &=& |A|^2 e^{-\Gamma t} 
[{1+|\xi |^2 \over 2} 
    - {1-|\xi |^2 \over 2} \cos{(\Delta M t)} + Im \xi \sin{(\Delta M t)}] , 
\label{IML} 
\end{eqnarray}  
where $\xi$ is defined as 
\begin{eqnarray} 
 \xi = \left ({q \over p}\right )_{B_d} {\bar A \over A}\;, 
\end{eqnarray}  
with $A = A(B^0 \rightarrow \pi^{+} \pi^{-})$ and 
$\bar A = \bar A(\bar B^0 \rightarrow \pi^{+} \pi^{-})$.  Here $p$ and $q$ are given by 

\begin{eqnarray} 
 |B_{L,H}> = p|B^0> \pm q|\bar B^0>\;,
\end{eqnarray}  
where $B_{H}$ and $B_{L}$ are the light and heavy mass eigenstates, 
respectively. 
In the SM $B^0-\bar B^0$ mixing
is dominated by the top quark in the loop, and therefore 

\begin{eqnarray}
\left ({q\over p}\right )_{ B_d} = {V_{tb}^*V_{td}\over V_{tb}V_{td}^*}\;.
\end{eqnarray} 

The decay amplitude can, in general, be parametrized as 
\begin{eqnarray}
\bar A_{\pi^+\pi^-} = V_{ub}V_{ud}^*T_{\pi^+\pi^-} + V_{tb}V_{td}^*P_{\pi^+\pi^-}\;,
\end{eqnarray}
where the amplitude $T$  contains both tree and penguin contributions, and $P$
contains penguin contribution only.
If the penguin amplitude $P_{\pi^+\pi^-}$ can be neglected, then 
\begin{eqnarray}
 Im \xi = Im ({V_{tb}^*V_{td}\over V_{tb}V_{td}^*} {V_{ub}V_{ud}^* \over
V_{ub}^*V_{ud}}) = \sin (2\alpha)\;.
\end{eqnarray}
The angle $\alpha$ can therefore be determined. 
However, if penguin effects are significant, the above method fails.

The decay $\bar B^0\rightarrow \pi^+\pi^-$ is 
induced by the effective Hamiltonian $H_{\Delta B =1}(d)$, and can 
be written as

\begin{eqnarray}
T_{\pi^+\pi^-} &=&
 {4 G_{F} \over \sqrt{2}} <\pi^+\pi^-|[
c_1 O^{u}_1(d) +c_2 O^{u}_2(d))+\sum_{i=3}^{10}(c_i^t-c_i^u)O_i(d)]|\bar B^0>\nonumber\\  
P_{\pi^+\pi^-} &=&{4G_F\over \sqrt{2}}
\sum_{i=3}^{12} <\pi^+\pi^-|(c_i^t-c_{i}^c ) O_{i}(d)|\bar B^0>\;.
\end{eqnarray}
Since the KM factors $|V_{ub}V_{ud}^*|$ is the same order of magnitude
compared with $|V_{cb}V_{cd}^*|$, the penguin contribution to the amplitudes
are at the level of a few percent compared with the tree amplitudes. 
However, even such a small contribution may cause significant error in the
determination of $\alpha$. It has been estimated~\cite{d-h2} 
that the error can be
as large as $12^o$. It is necessary to find ways to isolate the penguin
contributions.

When penguin effects are included, the parameter $Im\xi$ for 
$\bar B^0\rightarrow \pi^+\pi^-$ becomes~\cite{gl}
\begin{eqnarray}
Im \xi = {|\bar A|\over |A|}\sin(2\alpha + \theta)\;.
\end{eqnarray}

\begin{figure}[htb]
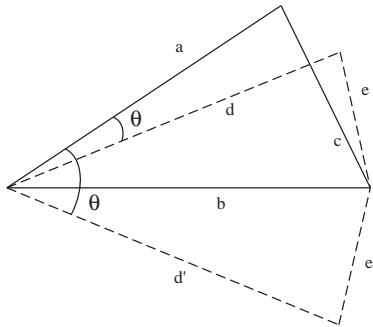

\centerline{ \DESepsf(IsoTriangles.epsf width 5 cm) }
\caption {Isospin triangles in the complex plane. Lines $a$, $b$, and $c$ denote
 the amplitudes
   $\bar A(B^0 \rightarrow \pi^+ \pi^-)$,
   $\protect \sqrt{2} \bar A(B^- \rightarrow \pi^- \pi^0) = \protect\sqrt{2}
A(B^+ \rightarrow \pi^+ \pi^0)$,
   and $\protect\sqrt{2} \bar A(B^0 \rightarrow \pi^0 \pi^0)$,respectively. The
dashed lines $d$ and $e$ (or $d'$ and
   $e'$) denote the amplitudes $A(B^0 \rightarrow \pi^+ \pi^-)$ and
   $\protect\sqrt{2} A(B^0 \rightarrow \pi^0 \pi^0)$, respectively.}
\label{isospin}
\end{figure}

To determine $\theta$, Gronau and London\cite{gl} proposed to use isospin relation
\begin{eqnarray}
\sqrt{2} \bar A(\bar B^0\rightarrow \pi^0\pi^0) +\sqrt{2} \bar A( B^-\rightarrow \pi^-\pi^0) = \bar 
A(\bar B^0\rightarrow \pi^+\pi^-)\;,
\end{eqnarray}
and similar relation for the $CP$-conjugate amplitudes for
the corresponding anti-particle decays. If all the six amplitudes can be measured, the angle 
$\theta$ can be determined up to two fold ambiguity as shown in Fig.\ref{isospin}.  
This is a very interesting theoretical idea. 
Experimentally, it may be difficult to measure $\theta$ accurately because
the branching ratio for $\bar B^0 \rightarrow \pi^0\pi^0$ is expected to be of
order $O(10^{-6})$. It has been pointed out that measurements for amplitude 
differences
in $B \rightarrow \pi K$ may help the measurements in $B\rightarrow \pi\pi$  
and improve the situation~\cite{desh-he}.

$\bar B^0 \rightarrow \pi \rho$ are induced by the same effective Hamiltonian. 
Similarly the 
penguin contamination can be removed by isospin analysis. These decay modes
also provide a measurement for $\alpha$~\cite{quin}. Combining this measurement with 
that from $B\rightarrow 
\pi \pi$, the two fold ambiguity mentioned above can
be eliminated.

\subsection{The determination of the angle $\beta$}

The best way to determine $\beta$ is to measure $Im\xi$ for
$\bar B^0 ( B^0) \rightarrow J/\psi K_S$~\cite{sanda}. 
The decay amplitude can be 
parameterized as
\begin{eqnarray}
A(\bar B^0\rightarrow J/\psi K_S) 
= V_{cb}V_{cs}^* T_{\psi K} + V_{ub}V_{us}^* P_{\psi K}\;.
\end{eqnarray}
The WC's involved 
indicate that $|T_{\psi K}|$ is much larger than $|P_{\psi K}|$.  Also 
$|V_{cb}V_{cs}^*|$ is about 
50 times larger than $|V_{ub}V_{us}^*|$  from experimental data.  
One can safely 
neglect the contribution from the term proportional to $V_{ub}V_{us}^*$. 
To a very good approximation, 
\begin{eqnarray} 
Im \xi = Im\left (\left ({q\over p}\right )_{B_d}{V_{cb}V_{cs}^*\over 
V_{cb}^*V_{cs}}\left ({q\over p}\right )_K \right ) = -\sin(2\beta)\;.
\end{eqnarray}
Here $(q/p)_K$ is the mixing parameter for $K^0-\bar K^0$ which is given
by $V_{cs}V_{cd}^*/V_{cs}^*V_{cd}^*$ and is small in the SM.
$\beta$ can be 
measured accurately.
This is the Gold-plated place to look for $CP$ violation.

\subsection{The determination of the angle $\gamma$}

The measurement of the angle $\gamma$ is an interest one. 
All methods proposed to measure $\gamma$ containing contributions 
from penguins involve additional 
assumptions about hadronic matrix elements
which are subject to further improvement~\cite{fles}. The best method to 
measure $\gamma$
is to use processes induced by the tree amplitudes for
$b \rightarrow u\bar c s$ and $b \rightarrow c \bar u s$. 

\begin{figure}[htb]
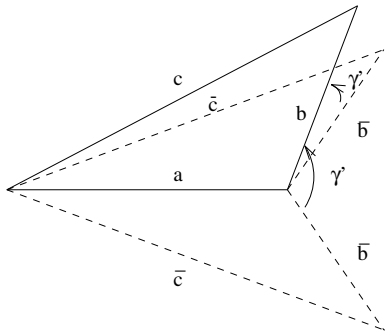

\centerline{ \DESepsf(Gamma.epsf width 5 cm) }
\caption {The measurement of $\gamma$ through $B^-(B^+)\rightarrow
D K^-(K^+)$ decays with $a = |\bar A(B^-\rightarrow D^0 K^-)| =
|A(B^+\rightarrow \bar D^0 K^+)|$, $b = \bar A(B^-\rightarrow \bar D^0)$,
$\bar b = A(B^+\rightarrow D^0 K^+)$, $c = \protect \sqrt{2} \bar A(B^-
\rightarrow D_{CP} K^-)$, and $\bar c = \protect \sqrt{2}
A(B^+\rightarrow D_{CP} K^+)$. }
\label{gamma}
\end{figure}

Let me give an
example based on the measurements of~\cite{wyler} 
$B^-\rightarrow (D^0,\; \bar D^0,$ $D_{CP}) K^-$. 
Here $D_{CP} = (D^0-\bar D^0)/\sqrt{2}$ is the CP even state.
The decay amplitudes can be parameterised as
\begin{eqnarray}
\bar A(\bar D^0 K^-) &=& V_{ub}V_{cs}^*T_{\bar D K}\;,\;\;
\bar A(D^0 K^-)=V_{cb}V_{us}^*T_{D^0K}\;,\nonumber\\
\bar A(D_{CP} K^-) &=& 
{1\over \sqrt{2}} (\bar A(D^0 K^-) - \bar A(\bar D^0 K^-))\;.
\end{eqnarray}
The angle $\gamma$ can be measured as shown in Fig.\ref{gamma}. 
The identification of $D_{CP}$ is through processes induced by $c\rightarrow
u d\bar d$ and $\bar c \rightarrow \bar u \bar d d$. The angle $\gamma'$
in Fig.\ref{gamma} 
is given by the absolute value of $\mbox{Arg}[(V_{ub}V_{cs}^*/V_{cb}
V_{us}^*)(V_{cd}V_{ud}^*/V_{cd}^*V_{ud})] = -2(\gamma-\sigma')$. In the SM
$\sigma'$ is very small, so $\gamma'$ is equal to 
$2\gamma$ to a very good approximation.
There is a two-fold ambiguity in the determination of $\gamma$ as shown in 
Fig.\ref{gamma}. This ambiguity
can be eliminated when combined with other measurements like,
$ B^- \rightarrow D K^{*-}$ and other similar decays~\cite{soni}.

\subsection{Other ways of testing the SM for $CP$ violation}

The error on the measurement of $|V_{ub}/V_{cb}|$ may well be under control. 
One can measure this ratio and two other phase angles to 
test the SM~\cite{d-b}. In fact this may be a  more convenient 
way to measure $CP$ violation in the KM sector in the presence of new source
for $CP$ violation which I will return to later.

It is very optimistic that the SM for $CP$ violation will be tested 
at $B$ factories. 

\section{Models For $CP$ Violation}

There are many other models for CP violation. In the following two
representative models will be discussed. One is $CP$ violation due to
spontaneous symmetry breaking~\cite{spon,weinberg}  and 
another $CP$ violation due to right-handed charged current 
in Left-Right symmetric model~\cite{l-r}. 
These two types of models
have interesting features. In the SM $CP$ is explicitly
violated. In 1973 T.D. Lee  first pointed out that $CP$ can 
actually be broken 
spontaneously~\cite{spon}. This opened a new direction in the study of 
$CP$ violation~\cite{cheng1}. 
Another interesting feature
of CP violation in the SM is that it only appears in the left-handed charged 
current. It is not sensitive to the right-handed sector. However, this is 
changed in the Left-Right symmetric model because the existence of right-handed
charged current~\cite{l-r}. 
$CP$ can be violated by phases in the right-handed charge current.
In these  models, it is not necessary to have at least
three generations to violate $CP$.

\subsection{Spontaneous $CP$ violation}

\begin{figure}[htb]
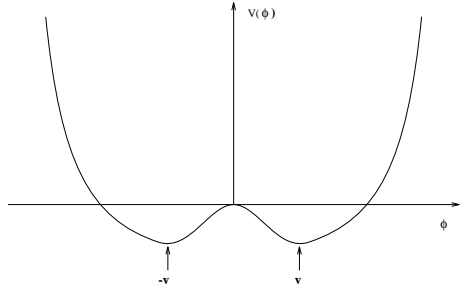

\centerline{ \DESepsf(fig-potential.epsf width 6 cm) }
\caption {
The potential for the $\phi$ field.} 
\label{potential}
\end{figure}

The basic idea for spontaneous $CP$ violation is best illustrated by 
using a toy
model given by T.D. Lee. The Lagrangian for this model is~\cite{spon} 

\begin{eqnarray}
L &=& -{1\over 2} \partial_\mu \phi \partial^\mu \phi - V(\phi)\nonumber\\
&-& {i\over 2} (\bar \psi \gamma_\mu \partial^\mu \psi -\partial^\mu \bar \psi
\gamma_\mu \psi) + m\bar \psi \psi - ig \bar \psi\gamma_5 \psi \phi\;,
\end{eqnarray}
where $\phi$ is a pseudoscalar field and $\psi$ is a spinor, $V(\phi)$ is
the potential for $\phi$ field. It is given by
\begin{eqnarray}
V(\phi) = {1\over 8} k^2(\phi^2-v^2)^2\;.
\end{eqnarray}
The transformation properties of $\phi$ under $P$, $C$ and $T$ are:
\begin{eqnarray}
P\phi = - \phi\;,\;\;C\phi = \phi\;,\;\; T\phi = -\phi\;.
\end{eqnarray}
The spinor has the usual transformation properties.
If the $\phi$ field does not develop any VEV, that is,
$<\phi> = 0$, the above
model is invariant under $CP$ transformation. However, the zero VEV 
for $\phi$ is not the minimal of the potential. The minimal occurs at
$<\phi> = \pm v$ as shown in Fig.\ref{potential}.
In the broken phase (in the phase with 
$<\phi>= v$, for example), the Lagrangian is given by,
\begin{eqnarray}
L &=& -{1\over 2} \partial_\mu H \partial^\mu H - V(H)\nonumber\\
&-&{i\over 2} (\bar \psi \gamma_\mu \partial^\mu\psi - \partial^\mu \bar \psi 
\gamma_\mu \psi) + m \bar \psi \psi - i gv\bar \psi \gamma_5 \psi - ig \bar 
\psi \gamma_5 \psi H\;,\nonumber\\
V(H) &=& {1\over 8} k^2 (H^2 +2vH)^2\;.
\end{eqnarray}
Here the field $H$ is defined as $\phi = v + H$ which has the same 
$C$, $P$ and $T$ 
transformation properties as $\phi$ in the unbroken phase. The VEV 
$v$ is a constant which does not transform under $P$, $C$ and $T$. 
The
potential $V(H)$ under $CP$ transformation becomes: 
$V^{CP}(H) = (1/8)k^2(H^2-2vH)^2$,
and the term $igv\bar \psi \gamma_5 \psi$ changes sign under $CP$. $CP$ is
spontaneously broken in the model. It is violated both in the scalar potential
and in the scalar-fermion interaction sectors. 

$CP$ violation in the 
scalar-fermion
interaction sector is more transparent if one works 
in the fermion mass eigenstate
basis 
where the mass is a 
real number. This can be achieved by a chiral rotation on the
fermion, $\psi'= exp(i\alpha \gamma_5/2) \psi$ with $tan(\alpha) = gv/m$. In
this basis, the kinetic energy term has the same form as in the 
$\psi$ basis, but
the mass and fermion-scalar interaction terms will be changed. One has

\begin{eqnarray}
L(\psi') = 
\sqrt{m^2 + g^2 v^2} \bar \psi' \psi' 
- g \bar \psi' (i\gamma_5 cos\alpha - sin\alpha) \psi' H + ...\;.
\end{eqnarray}

The $H$ field has both scalar and pseudoscalar couplings to
the fermion $\psi'$. Exchange of $H$ between fermions violates $CP$.

In the SM it is not possible to have spontaneous $CP$ violation. It
requires at least two Higgs doublets 
to have a realistic model.
With two Higgs doublets $H_1$ and $H_2$ transforming as $(1,2,-1)$ under the
SM gauge group, the most general Higgs potential one can write is~\cite{spon}:

\begin{eqnarray}
V(H_1, H_2) &=& \mu^2_i H^\dagger_i H_j + 
\lambda_i (H_i^\dagger H_i)^2 + \lambda'_i (H_i^\dagger H_j)(H_j^\dagger H_i)
\nonumber\\
&+& [\delta_1 (H^\dagger_1 H_2)(H^\dagger_1H_2)
+ \delta_2 (H^\dagger_1 H_2)(H^\dagger_1 H_1)\nonumber\\
&+& \delta_3 (H^\dagger_1 H_2)(H_2^\dagger H_2) + H.C. ]\;.
\end{eqnarray}
This potential only exhibits two possible electric charge conserving
minimal characterised by the VEV's,
$<H_i> = v_i exp(i\theta_i)$, and
classified according to the value of the relative angle $\theta = 
\theta_1-\theta_2$~\cite{cp-two}:

\begin{eqnarray}
\theta&=&0\;\;\;\;\;\mbox{if}\;\;\delta_1 < 0\;,\nonumber\\
\theta&=& arc cos \left ({\delta_2v_1^2 + \delta_3 v_2^2\over 4 \delta_1 v_1v_2}\right )\;\;\mbox{if}\;\; \delta_1 >0\;.
\end{eqnarray}
The solution with $\theta = 0$ does not violate $CP$, but the other one 
does.

Since $H_{1,2}$ have the same gauge transformation properties,
their couplings to quarks are similar. The most general Yukawa interactions
are given by

\begin{eqnarray}
L_{Y} = \bar Q_L (\lambda_1^U H_1 + \lambda_2^U H_2) U_R 
+ \bar Q_L (\lambda_1^D \tilde H_1 + \lambda_2^D \tilde H_2) D_R + H.C.
\end{eqnarray}

\begin{figure}[htb]
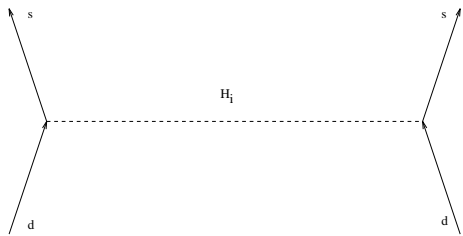

\centerline{ \DESepsf(fig-flavour.epsf width 6 cm) }
\caption {
Flavour changing neutral current at tree level.
}
\label{flavour}
\end{figure}

In this model there are three physical neutral Higgs and two charged
Higgs particles. 
In general the neutral Higgs particles have
flavour changing interactions at the tree level if $\lambda_1^{U,(D)}$ 
and
$\lambda_2^{U,(D)}$ are not proportional~\cite{cp-two,wolf-liu,hall-w,l-wu}. 
In fact there is no
reason why they should be proportional. These flavour changing neutral current
will induce $K^0- \bar K^0$ mixing at the tree level through the diagram shown 
in Fig.\ref{flavour}. The Higgs particles are constrained to be very 
heavy~\cite{cp-two,wolf-liu,l-wu}. 
There are rich phenomena related to $CP$ violation in this 
model~\cite{wolf-liu,l-wu} which will not be discussed any further.
 Instead I will 
consider models which do not have tree level flavour changing neutral currents.
This can be achieved if additional symmetries are imposed on the
model such that only 
one of the Higgs doublets couples to each of the up, down quarks and
the charged leptons. 

If the additional symmetry is imposed on the entire 
model with two Higgs doublets, 
the parameter $\delta_i$ must all be zero if the vacuum state of the Higgs potential 
is at the minimal. There is no 
solution for spontaneous $CP$ violation. In order to achieve spontaneous $CP$ 
violation, at least three Higgs doublets are needed.
A minimal model is the Weinberg model~\cite{weinberg}. In this model there are
three Higgs doublets $H_{1,2,3}$. 
The following is a set of possible discrete symmetries $D_1$ and $D_2$ 
which can achieve the goal~\cite{branco},

\begin{eqnarray}
&&D_1: Q_L \rightarrow Q_L\;,\;\;U_R\rightarrow U_R\;;\;\;D_R\rightarrow -D_R\;,
\;\;H_2\rightarrow -H_2\;,\;\;H_{1,3}\rightarrow H_{1,3}\;.\nonumber\\
&&D_2: Q_L\rightarrow Q_L\;,\;\;U_R\rightarrow U_R\;,\;\;D_R\rightarrow D_R\;,\;\;
H_{1,2}\rightarrow H_{1,2}\;,\;\;H_3\rightarrow -H_3\;.\nonumber\\
\end{eqnarray}

The Higgs potential is given by

\begin{eqnarray}
V(H_i) &=& \mu^2_i (H_i^\dagger H_i) + \delta_i (H^\dagger_i H_i)^2
+ \delta'_{ij}(H_i^\dagger H_j)(H^\dagger_j H_i)\nonumber\\
&+& \delta_{12} (H_1^\dagger H_2)(H^\dagger_1 H_2) +\delta_{13}
(H_1^\dagger H_3)(H_1^\dagger H_3)\nonumber\\
& +& \delta_{23} (H_2^\dagger H_3)(H_2^\dagger
H_3) + H.C.,
\end{eqnarray}
and the Yukawa interaction for quarks are given by

\begin{eqnarray}
L_Y = Q_L \lambda^U H_1 U_R + \bar Q_L \lambda^D \tilde H_2 D_R + H.C.
\end{eqnarray}

If all the constants in the Higgs potential and in the Yukawa interaction are 
real, there is no explicit $CP$ violation. After
symmetry breaking the situation is changed . The VEV's 
of the Higgs doublets can develop relative phases,
$<H_i> = exp(i\theta_i)$, if
$\lambda_{12} \lambda_{13}\lambda_{23} \ne 0$.
Minimising the Higgs potential one obtains
   
\begin{eqnarray}
C_{CP} = \lambda_{12}v_1v_2sin(\theta_2-\theta_1) 
= \lambda_{13}v_1v_3sin(\theta_3-\theta_1) = \lambda_{23} v_2v_3sin(\theta_3-\theta_2)
\;.
\end{eqnarray}
If $C_{CP}$ is not zero, $CP$ is violated.

In this model, 
there are five neutral Higgs and four charged Higgs 
particles. 
The couplings
of these Higgs particles to quarks can be written as

\begin{eqnarray}
L_{CC} &=& 2^{7/4} G_F^{1/2} \bar U [V_{KM} M_D ( \alpha_1 H_1^+ + \alpha_2 H_2^+)
R \nonumber\\
&+& M_U V_{KM} (\beta_1 H_1^+ + \beta_2 H_2^+)L]D + H.C.\nonumber\\
L_{NC} &=& 2^{7/4} G_F^{1/2} 
\sum_i m_i \bar q_i (\sigma_i + i \rho_i\gamma_5) q_i H_i^0\;,
\end{eqnarray}
where $q_i$ is one of the quarks. 
 $V_{KM}$ is real because there is no explicit violation of $CP$.
Because $C_{CP}$ is non-zero, $Im(\alpha_1\beta_1) = 
Im(\alpha_2 \beta_2)$ 
and $\sigma_i \rho_i$ are non-zero~\cite{tye}. These imply
$CP$ violation in the Yukawa interactions.

In later discussions it will be assumed that all Higgs
particles are heavy 
except one for each charged and neutral Higgs particles. They will be indicated 
by a subindex "1". 

In this model 
$CP$ is also violated in the lepton sector through Yukawa couplings. 
There are different ways leptons can couple to Higgs particles. 
One of the possibilities is 
to assume that $H_3$ is the Higgs which 
couples to leptons. In this case the leptons
transform under the discrete symmetry as 

\begin{eqnarray}
D_1: && L_L \rightarrow L_L\;,\;\;E_R\rightarrow E_R\;;\nonumber\\
D_2:&&L_L\rightarrow L_L\;,\;\; E_R \rightarrow -E_R\;.
\end{eqnarray}

\begin{figure}[htb]
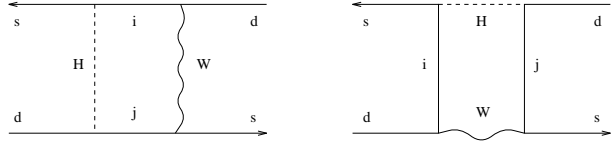

\centerline{ \DESepsf(fig-box-wein.epsf width 8 cm) }
\caption {
The new "box" diagrams in the Weinberg model.}
\label{box-w}
\end{figure}

\begin{figure}[htb]
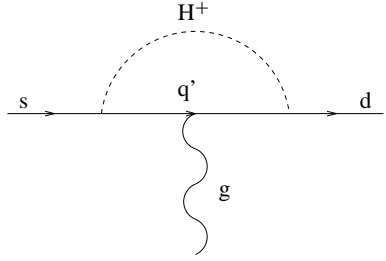

\centerline{ \DESepsf(fig-higgs-dipole.epsf width 5 cm) }
\caption {
The strong dipole penguin diagram due to Higgs exchange in the Weinberg model.}
\label{higgs-dipole}
\end{figure}

The Weinberg 
model can easily explain the observed $CP$ violation in $K^0-\bar K^0$
mixing. Naively, one would expect that the $CP$ violating parameter $\epsilon$
is induced by the "box" diagrams as shown in Fig. \ref{box-w}. 
This turned out to be problematic 
because the enhanced $\Delta S =1$ interaction, i.e., the 
gluon dipole penguin interaction shown in Fig.\ref{higgs-dipole},
\begin{eqnarray}
L_{CP} &=& 
i{G_F\over \sqrt{2}}{g_s\over 32\pi^2} m_s \sum_{ij}x_{ij}
(V_{id}^*V_{is}) 
\bar f_{ij} \bar d \sigma_{\mu\nu} (1+\gamma_5) \lambda^a s G_a^{\mu\nu}\;,
\nonumber\\
\bar f_{ij} &=& -
Im(\alpha_i\beta^*_i)
[{1\over 2(1-x_{ij})} + {1\over (1-x_{ij})^2} + {1\over (1-x_{ij})^3}
lnx_{ij}]\;,
\label{dipole}
\end{eqnarray}
where $x_{ij} = m_j^2/m_{H_i}$.

\begin{figure}[htb]
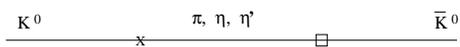

\centerline{ \DESepsf(fig-epsilon-wein.epsf width 6 cm) }
\caption {
The dominant contribution to $\epsilon$ in the Weinberg model.
The cross indicates a $CP$ violating weak vertex and the 
square indicates a $CP$ 
conserving weak vertex.}
\label{epsilon-ew}
\end{figure}

If 
$\epsilon$ is purely due to the "box" diagrams in Fig.\ref{box-w}, 
the magnitude for
the $CP$ violating parameter $\epsilon'/\epsilon$ is predicted to be too large
compared with experimental data~\cite{s-d}.
However, it was pointed out~\cite{dhg,cheng} that if $\epsilon$ is actually
dominated by the long distance interaction as shown in Fig.\ref{epsilon-ew}, the
problem can be solved. 
The $CP$ violating part of $M_{12}$ is given by~\cite{cheng},

\begin{eqnarray}
Im(M_{12}) &=& {1\over 4m_K} \sum_{i=\pi,\eta,\eta'}
{Im(<K^0|L|i><i|L|\bar K^0>)\over m_K^2 - m_i^2}\nonumber\\
&=&{2\kappa\over m_K^2-m_\pi^2} <K^0|L_{CP}|\pi^0><\pi^0|L|\bar K^0>\;.\nonumber\\
\kappa&=& 1 + {m_K^2-m_\pi^2\over m_K^2-m_\eta^2} (\sqrt{{1\over 3}}(1+\delta)
cos\theta + 2 \sqrt{{2\over 3}}\rho sin\theta)^2\nonumber\\
&+& {m_K^2-m_\pi^2\over m_K^2-m_{\eta'}^2} (\sqrt{{1\over 3}}sin\theta - 2 
\sqrt{{2\over 3}} \rho cos\theta)^2\;,
\end{eqnarray}
where~\cite{chau-cheng} 
$\theta \approx -20^o$ is the $\eta - \eta'$ mixing angle, $\delta = 0.17$
and $\rho = 0.75\pm 0.05$ parameterize flavour symmetry breakings,

\begin{eqnarray}
<\eta_8|L|K^0> &=& \sqrt{{1\over 3}} (1+\delta) <\pi^0|L|K^0>\;,\nonumber\\
<\eta_0|L|K^0> &=& -2\sqrt{{2\over 3}}\rho <\pi^0|L|K^0>\;.
\end{eqnarray}

Using the facts that due to mass suppression factor for the 
$u$ quark in the loop and KM suppression factor for $t$ in the loop, the
dominant contribution is from $c$ quark in the loop~\cite{cheng}, one 
obtains

\begin{eqnarray}
4m_K Im(M_{12}) = -4.3\times 10^{-16} \bar f_{1c}\;GeV^2\;,
\label{epsilon-w}
\end{eqnarray}
and the function $\bar f_{1c}$ is determined to be 
$(0.054 \sim 0.061)$ GeV$^{-2}$, a reasonable value to have.

At this point I would like to point out that if one abandons
the requirement of spontaneous $CP$ violation, 
it is possible to have $CP$ violation in both 
the KM and the Higgs sectors. Such
models will have more flexibility to accommodate experimental data.

\subsection{ Left-Right Symmetric Model}

The gauge group of the Left-Right symmetric model is
$SU(3)_C\times SU(2)_L\times SU(2)_R\times U(1)_{B-L}$~\cite{l-r,ll-r}. 
This model offers
a possible explanation of why the observed weak charged current 
interactions are all left-handed, but not right-handed.
In this model there are both left-handed and right-handed charged currents. 
The 
answer to the  question asked is that spontaneous symmetry breaking first breaks the
$SU(2)_R\times U(1)_{B-L}$ at a higher energy scale 
to $SU(3)_C\times SU(2)_L\times U(1)_Y$. Because the interaction strength 
is inversely proportional to the square of the energy scale,
the right-handed current effects are suppressed. However, the appearance of 
right-handed current introduces non-negligible 
effects for $CP$ violation. 

In Left-Right symmetric model the right-handed fermions are grouped into doublets under
$SU(2)_R$. For the leptons, this requires the introduction of right-handed
neutrino. The gauge group transformation properties of the fermions are:

\begin{eqnarray}
Q_L: (3,2,1,1/3)\;,\;\;&& Q_R:(3,1,2,1/3)\;,\nonumber\\
L_L: (1,2,1,-1)\;,\;\;&&L_R:(1,1,2,-1)\;.
\end{eqnarray}

Neutrinos are massive in this model. Depending on whether neutrinos have
only dirac masses or have both dirac and majorana masses, the separation of the breaking scales
for $SU(2)_R$ and $SU(2)_L$ can be achieved differently.

If neutrinos have dirac masses only, the desired symmetry breaking pattern 
can be achieved by introducing~\cite{ll-r} 
\begin{eqnarray}
H_L:(1,2,1,1)\;,\;\;H_R:(1,1,2,1)\;,
\end{eqnarray}
with $<H_R> = v_R$ to be much larger than $<H_L> = v_L$.
Therefore
$SU(2)_R$ is broken at a larger energy scale than 
the one for  $SU(2)_L$ breaking.
Whereas if 
neutrinos have both dirac and mjorana masses, the desired symmetry 
breaking can be achieved by~\cite{ll-r}

\begin{eqnarray}
\Delta_L:(1,3,1,2)\;,\;\;\Delta_R:(1,1,3,2)\;,
\end{eqnarray}
with 
$<\Delta_R>= v_R$ much larger than $<\Delta_L> = v_L$.

To generate fermion masses, it is necessary to introduce Higgs bi-doublet

\begin{eqnarray}
\phi: (1,2,2,0) = \left (\begin{array}{ll}
\phi_1^0&\phi_1^+\\
\phi_2^-&\phi_2^0
\end{array}
\right )\;.
\end{eqnarray}

There are left-handed as well as right-handed charged currents in this model. 
They are

\begin{eqnarray}
L_{CC} = {1\over \sqrt{2}}(g_L \bar U \gamma^\mu L D W^+_{L\mu} + 
g_R \bar U \gamma^\mu R D W^+_{R\mu})\;,
\end{eqnarray}
where $g_L$ and $g_R$ are the gauge couplings for $SU(2)_L$ and $SU(2)_R$,
respectively.

Because the VEV's of $\phi$ break both $SU(2)_R$ and
$SU(2)_L$, there is mixing between $W_R$ and $W_L$. The mass eigenstates
$W_1$ and $W_2$ are related to the weak eigenstates by

\begin{eqnarray}
\left ( \begin{array}{l}
W_1\\
W_2
\end{array}
\right ) = \left ( \begin{array}{ll}
cos\xi&\sin\xi\\
-sin\xi&cos\xi
\end{array} 
\right )
\left (\begin{array}{l}
W_L\\
W_R
\end{array}
\right )\;.
\end{eqnarray}

Writing the charged current interactions in
the gauge boson mass eigenstate
 as well as the quark mass eigenstate basis, the charged currents become, 

\begin{eqnarray}
L_{CC} 
&&= {1\over \sqrt{2}} ( g_L \bar U V_{KM}^L \gamma_\mu L D cos\xi 
+ g_R \bar U V_{KM}^R\gamma_\mu R D sin\xi) W_1^\mu\nonumber\\
&&+ {1\over \sqrt{2}}(g_R \bar U V_{KM}^R \gamma^\mu R D cos\xi 
-g_L \bar U V_{KM}^L \gamma_\mu L D sin\xi)W_2^\mu
+ H.C.
\end{eqnarray}
where $V_{KM}^{L,R}$ are the equivalent $KM$ matrices 
for the left-handed and right-handed
charged currents. Just like in the SM one can always 
absorb $2N-1$ parameters in the $KM$ matrix by redefining quark phases,
one can choose a basis such that $V_{KM}^L$ is the
same as in the SM. However, after this choice is made, 
there is no freedom to absorb
parameters in $V_{KM}^R$. There are $N(N+1)/2$ phases in $V_{KM}^R$. $CP$
symmetry can be violated even with just one generation. 

The observed $CP$ violation in $K^0-\bar K^0$ mixing can be easily 
accommodated. An interesting scenario is that mxings only occur among the
first two generations.
In this case $V_{KM}^L$ is real. $CP$ violating 
phases only exist in $V_{KM}^R$. 
There are three $CP$ violating phases appearing in $V_{ij}^R$, 
it can be parameterized as

\begin{eqnarray}
V_{KM}^R = e^{i\gamma}\left (\begin{array}{ll}
e^{-i\delta_2}cos\theta^R &e^{-i\delta_1}sin\theta^R \\
-e^{i\delta_1}sin\theta^R&e^{i\delta_2}cos\theta^R
\end{array}
\right )\;.
\end{eqnarray}

\begin{figure}[htb]
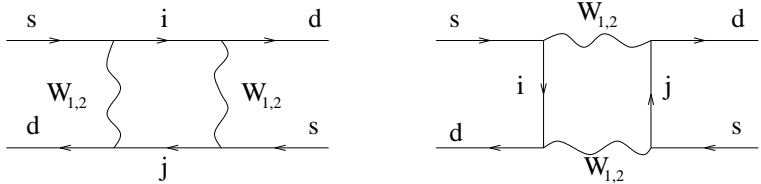

\centerline{ \DESepsf(fig-epsilon-lr.epsf width 10 cm) }
\caption {
The dominant contribution to $CP$ violation in $K^0-\bar K^0$ mixing 
to $\epsilon$ in the Left-Right symmetric model with two generation mixing.}
\label{epsilon-lr}
\end{figure}

Because there is no $CP$ violation in the purely left-handed current
interaction, right-handed 
charged current must provide the needed source for $CP$ violation.
The dominant contribution is from the diagrams shown in Fig.\ref{epsilon-lr}.
Assuming $g_L = g_R$ and $|V{ij}^L| = |V_{ij}^R|$, one obtains

\begin{eqnarray}
\epsilon = {1\over 2\sqrt{2}} 
430 {m_{W_1}^2\over m^2_{W_2}} sin(\delta_2-\delta_1)e^{i\pi/4}\;.
\end{eqnarray}

There is constraints on the right-handed gauge boson mass from the mass 
difference between $K_S$ and $K_L$, $(m_{W_1}/m_{W_2})^2 
<1/430$~\cite{soni-l,chang}. 
This implies that the $CP$ violating parameter $|sin(\delta_2-\delta_1)|$
must be larger than  
$2\sqrt{2} |\epsilon|$~\cite{soni-l,chang,hmp1}.

\section{The KM Unitarity Triangle And New Physics }

In this section I discuss ways to extract variables in the 
KM matrix in the presence of new $CP$ violating sources.
If new sources  exist, such as in the Weinberg and Left-Rgiht 
symmetric models, the interpretation 
of the measurements discussed in Section 3 will have to be 
modified~\cite{d-b,beyond,he-w}.

In general new CP violating interactions come in all possible ways. They
can arise at the tree and/or loop levels. For a certain 
process there may be contributions from several different CP violating
sources. It is important to isolate these sources. This is, of course,  a very
difficult task. 
In this section the
possibility to achieve this task by using the $B$ decay modes
discussed in Section 3, will be discussed.

If new CP violating interactions come at all stages, tree and loop, significantly,
it is possible to see the deviation from the SM, but it is not possible 
to
isolate individual contribution. However, in many models new CP violating
contributions only have significant effects at loop levels in $B$ decays, like
the Weinberg model. 
In the following I will 
concentrate on this class of models.

It has been point out in Section 3 that
it is possible
to determine the KM triangle by using just tree level processes, namely,
$\gamma$ from 
$B^- \rightarrow D K^-$ and  $|V_{ub}/V_{cb}|$ from $b \rightarrow u(c) l 
\bar \nu$. These two quantities will determine the shape of the unitarity 
triangle. 
>From these measurements, one knows for certain that if KM
mechanism is, at least partially,  responsible for CP violation. After this is
done one  can use the other processes to see what the new contributions
are.

Quantities generated at loop
level will have new $CP$ violating phases. The mixing parameters
$(q/p)_i$ for $B^0-\bar B^0$, and $K^0 -\bar K^0$ will be modified.
They can be normalised to the SM ones as the following
\begin{eqnarray}
\left ({q\over p}\right )_{B_d} = \left ({V_{tb}^*V_{td}\over V_{tb}V_{td}^*}\right )
e^{2i\delta_{B_d}}\;,\;\;
\left ({q\over p}\right )_K = \left ({V_{cd}^*V_{cs}\over V_{cd}V_{cs}^*}\right )
e^{2i\delta_K}\;.
\end{eqnarray}
Because of the smallness of $\epsilon$, $\delta_K$ is negligibly small 
($<10^{-3}$). Its contribution will be neglected.

The $B$ decay amplitudes involving loop corrections will also have new
phases. They can be written in the following form

\begin{eqnarray}
\bar A(\bar B^0\rightarrow \pi\pi) &=& V_{ub}V_{ud}^*e^{i\theta_{\pi\pi}} 
T_{\pi\pi}(new)\;, \nonumber\\
\bar A(\bar B^0\rightarrow \psi K_S) &=& V_{cb}V_{cs}^*e^{i\theta_{\psi K}} 
T_{\psi K}(new)\;.
\label{theta}
\end{eqnarray}

The measurements of $Im\xi$ for these processes 
 will no longer have the clean interpretation as 
in the SM. 
One has
\begin{eqnarray}
Im\xi_{\pi \pi} &=& Im\left ( \left ( {q\over p}\right )_{B_d} 
{\bar A(\pi \pi) \over A(\pi \pi)}\right )\nonumber\\
&=&\mbox{sin}[2(\alpha+\delta_B +\theta_{\pi\pi})]\;,\nonumber\\
Im\xi_{\psi K} &=& Im\left ( \left ( {q\over p}\right )_{B_d} \left (
{q\over p}\right )_K {\bar A(\psi K_S) \over A(\psi K_S)}\right )\nonumber\\
&=& -\mbox{sin}[2(\beta-\delta_B-\theta_{\psi K})]\;.
\end{eqnarray}
If one also assumes 
that the loop contribution only has substantial contribution to
$\delta_B$ and use $B \rightarrow \pi\pi$, $B\rightarrow J/\psi K_S$
and $B\rightarrow K^-( D^0,\;\bar D^0,\;D_{CP})$ as tests for the SM, one 
would still obtain the summation of the angles measured to be 
$180^o$ because $B\rightarrow \pi\pi$ would measure $\alpha+\delta_B$, 
$B\rightarrow J/\psi K_S$ would measure $\beta-\delta_B$ and $
B\rightarrow K^-(D^0,\;\bar D^0,\;D_{CP})$ would still measure $\gamma$. One 
would not be able to know if new physics has shown up. However, if one 
supplements the measurement of $V_{ub}/V_{cb}$, and use
the measurements of $\gamma$ and $V_{ub}/V_{cb}$ to fix the KM unitarity
triangle first and the additional measurements from $B\rightarrow \pi\pi$ and
$B\rightarrow J/\psi K_S$ will provide information about new physics.

Let me consider the Weinberg model in more detail~\cite{he-w} and
assume that $CP$ violation appears
in both the KM and Higgs sectors simultaneously.
The decay amplitudes 
due to exchange of charged Higgs at tree level will be proportional to 
$V_{fb}V_{f'q}^*(m_b m_{f'}/m_{H_i}^2)\alpha_i \beta_i^*$. Therefore if a
decay involves light quark,  the amplitude will be suppressed. Similar arguments
apply to semi-leptonic decays of b quark. It is clear that the measurement
of $\gamma$ and $|V_{ub}/V_{cb}|$ will not be affected.

 However, at one 
loop level if the internal quark masses are large, sizeable CP violating
decay amplitude may be generated. The leading term is from the strong dipole
penguin interaction, similar to the diagram in Fig.\ref{higgs-dipole} 
with top quark in the loop\cite{dhg1},
\begin{eqnarray}
L_{DP} &&= V_{tb}V_{tq}^*\tilde f O_{11}\;,\nonumber\\
\tilde f &&= {G_F\over 16\sqrt{2}}\sum_i^2 \alpha^*_i\beta_i
{m_t^2\over m_{H_i}^2-m_t^2}[
{m_{H_i}^4\over (m_{H_i}^2-m_t^2)^2} \mbox{ln}{m_{H_i}^2\over m_t^2}
-{m_{H_i}^2\over m_{H_i}^2-m_t^2} - {1\over 2}]\;,\nonumber\\
O_{11} &&= {g_s\over 32\pi^2} m_b \bar s \sigma_{\mu\nu}RT^ab G^{\mu\nu}_a\;.
\end{eqnarray}
This is not suppressed compared with the penguin contributions in the SM. 
There is also a similar contribution 
from the operator $O_{12}$. However the WC of this operator is 
suppressed by a factor of $\alpha_{em}/\alpha_s$ 
and its contribution can be neglected.
The contribution from $O_{11}$ can be written as

\begin{eqnarray}
\bar A_{final }(weinberg) = -(V_{ub}V_{uq}^*+
V_{cb}V_{cq}^*)a_{final}e^{i\alpha_H}\;,
\end{eqnarray}
where $\alpha_H$ is the phase in $\tilde f$ which is decay mode independent,
and $a_{final} = |\tilde f|<final \;state| O_{11}|B>$ which is
decay mode dependent. Due to this contribution, the phases $\theta_i$
defined in eq.(\ref{theta}) will not be zero. 

The charged Higgs bosons also contribute to the mixing parameters $(q/p)_i$.
These contributions have the same KM factors as the SM, but in principle
have additional phase factors due to non-zero $Im(\alpha_1\beta^*_1)$.

>From the above discussions, one sees that even with 
new contributions to $B$ decay amplitudes,
the measurements of $\gamma$ and $|V_{ub}/V_{cb}|$ using
$B^-\rightarrow (D^0,\;\bar D^0,\;D_{CP})K^-$
and $b\rightarrow u(c) l\bar \nu$ 
will be true measurements of these quantities.
CP violation due to KM mechanism can
be isolated.
It is not possible to use these two measurements to distinguish the
SM and the Weinberg model. 
However, if one also measures $Im \xi$ for $B\rightarrow \pi\pi$ and 
$\bar B^0
\rightarrow \psi K_S$, these two models can be distinguished because if the
SM is correct, the angles $\alpha$ and  $\beta$ are measured, whereas if the 
Weinberg model is correct,
the quantities $\alpha+\delta_B$ and $\beta-\delta_B-\theta_{\psi K}$ are 
measured.

The Left-Right symmetric model will have completely different results.
In two generation mixing case, there is no unitarity triangle to talk about. 
With three generations, even thought the left-handed current still has 
a unitarity triangle as in the SM one, 
due to the appearance of right-handed current, new physics will
come significantly at both the tree and loop levels. In general it is not 
possible to isolate the left-handed current contribution.

\section{Direct $CP$ Violation In Neutral Kaon System}
 
There are many other experiments which can test $CP$ violation in kaon system.
 I now discuss  direct $CP$ violation in $K\rightarrow \pi\pi$ decays. 
For this purpose, it is convenient to study the 
quantities $\eta_{+-}$ and $\eta_{00}$  
defined as the following~\cite{w-yang}

\begin{eqnarray}
\eta_{+-} = { A(K_L\rightarrow \pi^+\pi^-)\over A(K_S\rightarrow \pi^+\pi^-)}\;,
\;\;\eta_{00} = {A(K_L\rightarrow \pi^0\pi^0)\over A(K_S\rightarrow \pi^0\pi^0)}\;.
\end{eqnarray}
One can express the above quantities in terms of $K^0(\bar K^0)\rightarrow 
\pi\pi$ isospin decay amplitudes,

\begin{eqnarray}
A(K^0\rightarrow \pi^+\pi^-)&=& \sqrt{{2\over 3}} A_0 e^{i\delta_0}
+ \sqrt{{1\over 3}} A_2 e^{i\delta_2}\;,\nonumber\\
A(K^0\rightarrow \pi^0\pi^0)&=& \sqrt{{1\over 3}} A_0 e^{i\delta_0}
- \sqrt{{2\over 3}}A_2 e^{i\delta_2}\;,
\end{eqnarray}
where $A_0$ and $A_2$ are decay amplitudes for isospin $I= 0$ and $I=2$ 
final two pion systems, respectively. 
$\delta_i$ are the strong final state rescattering
phases (strong phase). 

The corresponding anti-particle decay amplitudes are

\begin{eqnarray}
A(\bar K^0\rightarrow \pi^+\pi^-)&=&- \sqrt{{2\over 3}} A_0^* e^{i\delta_0}
- \sqrt{{1\over 3}} A_2^* e^{i\delta_2}\;,\nonumber\\
A(\bar K^0\rightarrow \pi^0\pi^0)&=& -\sqrt{{1\over 3}} A_0^* e^{i\delta_0}
+ \sqrt{{2\over 3}}A_2^* e^{i\delta_2}\;.
\end{eqnarray}

One has

\begin{eqnarray}
\eta_{+-} &=& \epsilon + i{ImA_0\over ReA_0} + 
e^{i(\pi/2 +\delta_2-\delta_0)} {ReA_2\over \sqrt{2} A_2} \left (
{ImA_2\over Re A_2} - {ImA_0\over ReA_0}\right )\;,\nonumber\\
\eta_{00}&=& \epsilon + i {ImA_0\over ReA_0} 
-2 e^{i(\pi/2+\delta_2-\delta_0)} {ReA_2\over \sqrt{2} ReA_0}
\left ( {ImA_2\over ReA_2} - {ImA_0\over A_0}\right )\;,
\end{eqnarray}
and the parameter $\epsilon'$ mentioned previously is defined as

\begin{eqnarray}
\epsilon' = {\eta_{+-} - \eta_{00} \over 3} = {ReA_2\over \sqrt{2} ReA_0}
\left ( {ImA_2\over ReA_2} - {ImA_0\over ReA_0} \right )\;.
\end{eqnarray}

The strong phases $\delta_i$ can be determined from phase shift 
analyses in $\pi - \pi$ scattering, and $\pi/2 +\delta_2 - \delta_0$ is
found to be close to $\pi/4$. The requirement of $CPT$ symmetry implies that
this phase is equal to the phase $\phi_\epsilon$ 
for $\epsilon$. In the literature 
the quantity $\epsilon'/\epsilon$ is usually used.

Experimental measurement of this quantity is not conclusive. While the
result of NA31 at CERN~\cite{na31} with $\epsilon'/\epsilon
 = (23\pm 7)\times 10^{-4}$
clearly indicates direct $CP$ violation, the value of E731 at 
Fermilab~\cite{e731},
$\epsilon'/\epsilon = (7.4\pm 5.9)\times 10^{-4}$ is compatible with $CP$
conservation.

The measurement of $\epsilon'/\epsilon$ provides important information in
distinguishing superweak model and other models because superweak model 
predicts zero value for $\epsilon'/\epsilon$. 
This measurement
also provides constraints for other models. 
In the SM a non-zero value for $\epsilon'$ is generated at one loop level 
similar to the diagrams for $B$ decays as shown in Fig.\ref{decay}. This 
has been studied
extensively in the liturature~\cite{dhg2,randel,l-wu1,buras-e}. 
One feature particularly interesting 
is
that both the strong and electroweak penguin effects are 
important~\cite{dhg2,randel}. Without
electroweak penguin contribution, $\epsilon'/\epsilon$ is predicted to 
be larger than the experimental limit.
When
the electroweak penguin effect is included, 
the situation is changed because although
electroweak penguin contribution to the $I = 0$ amplitude is small, it
contributes to $I =2$ amplitude with substantial value for $ImA_2/ReA_2$. 
This new contribution tends to cancel 
$\epsilon'/\epsilon$ from the strong penguin. 
The final value is predicted to be
in the range of~\cite{buras-e} $(-1.2\sim 44)\times 10^{-4}$ 
which is consistent with present experimental limit.

\begin{figure}[htb]
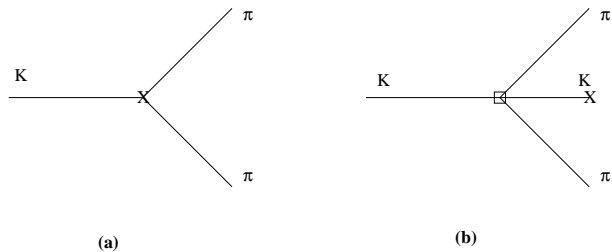

\centerline{ \DESepsf(fig-epsilon-prime.epsf width 8 cm) }
\caption {
Cancellation between diagrams in the Weinberg model. The cross indicates a weak 
vertex and the square indicates a strong vertex.}
\label{epsilon-prime}
\end{figure}

In the spontaneous $CP$ violation model, the most significant operator
contributing to $\epsilon'$ is from eq.(\ref{dipole}). 
If the experimental value for $\epsilon$
is purely from the "box" diagram shown is Fig.\ref{box-w}, 
the predicted value for
$\epsilon'/\epsilon$ due to contribution from eq.(\ref{dipole}) 
will be much large than the experimental
limit. This problem is solved if 
the long distance contribution dominates $\epsilon$  
as discussed earlier. If one naively use the
chiral realisation of $L_{CP}$ in eq.(\ref{dipole}) 
for $K \rightarrow \pi\pi$ and take
the direct diagram (a) in Fig.\ref{epsilon-prime}, 
one would obtain a large $\epsilon'/\epsilon$ which is in conflict with 
experimental data. However, it was pointed out that at the
same order there is another diagram (b) in Fig.\ref{epsilon-prime} 
which cancels the contribution from (a) in Fig.\ref{epsilon-prime}~\cite{dhg}. 
The contribution for $\epsilon'/\epsilon$ comes at higher
order and is suppressed.
The parameter $\epsilon'/\epsilon$ is predicted to 
be~\cite{cheng}
\begin{eqnarray}
{\epsilon'\over \epsilon} \approx 0.017 D\;,
\end{eqnarray}
with $D = O((m^2_\pi\; \mbox{or}\; m^2_K)/\Lambda^2_\chi)$ 
being a chiral suppression factor
which characterises cancellations discussed above. 
The value for $\epsilon'/ \epsilon$ is in the range of
$(0.4 \sim 6.0)\times 10^{-3}$ which is, again, 
consistent with present experimental limit.

In the Left-Right symmetric model with two generation mixing, 
the dominant contribution to $\epsilon'/\epsilon$ 
is due to $W_L-W_R$ mixing, one
obtains~\cite{hmp1}

\begin{eqnarray}
{\epsilon'\over \epsilon} = 276 \mbox{tg} \xi [sin(\gamma-\delta_2) + 
sin(\gamma-\delta_1) - 0.1(sin(\gamma+\delta_2)+sin(\gamma + \delta_1))]\;,
\end{eqnarray}
which can easily accommodate experimental data.

New experiment at DA$\phi$NE will
improve the measurement for $\epsilon'/\epsilon$ considerably~\cite{dane}. 
Models for $CP$ violation will be further constrained.

There are many other experiments studying $CP$ violation in kaon systems. They
have been discussed in several excellent reviews~\cite{kaon-cp}. 
I will not discuss them
here. In the following sections I will discuss $CP$ violation in other systems.

\section{The Electric Dipole Moment}

The interaction potential of an electric dipole 
$\vec D$ in an external
electric field $\vec E$ is proportional to $\vec D\cdot \vec E$.  
Classically the electric dipole moment (EDM) is given
by $\vec D = \int d^3x \vec x \rho(\vec x)$, where $\rho(\vec x)$ is the
electric charge density. In the case of an elementary particle, the only
(pseudo) vector that characterises its state is the spin $\vec S$; hence
$\vec D$ must be of the form $d \vec S$.
Here $d$ is a proportional 
constant representing the size of the EDM. Under $P$ transformation
$\vec D\rightarrow \vec D$, $\vec E\rightarrow -\vec E$, while 
under $T$ transformation, $\vec D\rightarrow -\vec D$ and $\vec E\rightarrow 
\vec E$. So the interaction $\vec D\cdot \vec E$ changes sign under $P$ and 
$T$ transformation. If $d$ is not zero, $P$ and $T$ are violated simultaneously.
This is a direct test
of time reversal symmetry. Due to the $CPT$ theorem, a non-zero value for $d$ 
also violates $CP$. The EDM of an elementary particle is of interest for
both experimental~\cite{p-r,d-exp,e-edm,atom-exp} and  
theoretical~\cite{hmp2,b-m,b-s,other-edm} studies.

\subsection{The neutron EDM}

The neutron EDM has been of interest to 
physicists for a long time. The measurement of the neutron 
EDM started in 1950 
by Purcel and Ramsey~\cite{p-r}.
Although no positive result has been obtained, 
very impressive progress (several orders of magnitude) 
on the upper bound has been obtained. 
The present experimental upper bound for the
neutron EDM $d$ is~\cite{d-exp} $1.2\times 10^{-25}$ ecm.
There are also many experiments measuring the EDM's of
other particle systems, like the electron~\cite{e-edm} 
and atoms~\cite{atom-exp}. Stringent bounds have
also been obtained~\cite{other-edm}.

There are different contributions to the neutron EDM. It can arise at the
hadron level as well as at the quark level.
At the quark level, it can come from the 
quark EDM $d_q$,  the color dipole moment (CDM) 
$f_q$, and other complicated $CP$ violating
operators composed of quarks and/or gluons. In the valance quark model,
the contributions from $d_q$ and $f_q$ are given by

\begin{eqnarray}
d(d_q) &=& {1\over 3}(4d_d - d_u)\;,\;\;d(f_q) = {1\over 9}e(4f_d + 3f_u)\;.
\end{eqnarray}

For complicated operators it is very difficult, if not impossible, 
to calculate their contributions to the neutron EDM. In this case dimensional
analysis may help to make an order of magnitude estimate. A commonly used 
nethod is the  so called "naive dimensional analysis" (NDA)~\cite{nda} 
which keeps track 
of factors of $4\pi$ from loops and mass scales involved.
Giving a $CP$ violating operator, $O$ with coefficient $C$, 
one defines the reduced 
coupling constant $(4\pi)^{2-N}\Lambda_\chi^{D-4}C$ where $\Lambda_\chi 
= 2\pi f_\pi = 1190 MeV$ 
is the chiral symmetry breaking scale, $N$ is the number of fields and $D$ is 
the dimension of the operator.  The neutron EDM operator has a reduced coupling 
$d_n \Lambda_\chi/4\pi$.  
For a $CP$ violating operator $O$ which does not involve photon, 
in order to produce a  neutron EDM a photon has to be attached to a quark.  
This electromagnetic coupling of quark has a reduced coupling $e/4\pi$.  
So the NDA suggests that the neutron EDM due to the operator $O$ is given by,
\begin{eqnarray}
d_n \approx {e\over \Lambda_\chi}(4\pi)^{2-N}\Lambda_\chi^{D-4}C.
\end{eqnarray}
Of course, one must keep in mind that 
this is only an order of magnitude estimate.

\begin{figure}[htb]
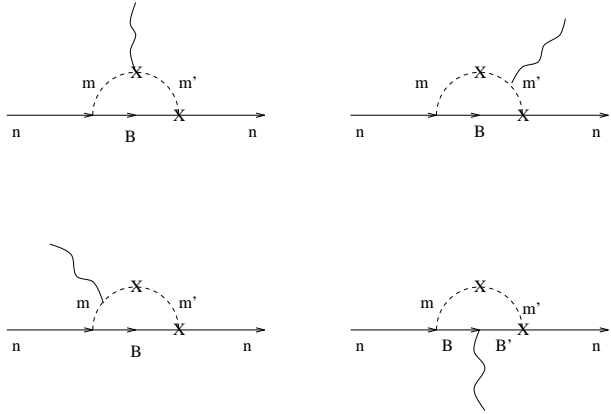

\centerline{ \DESepsf(fig-edm-sm.epsf width 8 cm) }
\caption {
Some typical diagrams 
generating the neutron EDM in the SM. Each cross indicates a
weak vertex which is equivalent to a loop contribution.}
\label{edm-sm}
\end{figure}

It has been shown that in the SM the quark 
EDMs are
zero at one and two loop levels~\cite{sha}. The neutron
EDM is generated at three loop level and therefore
is very small in size. A typical set of diagrams 
which generate  neutron EDM with hadron loops is shown in
Fig.\ref{edm-sm}~\cite{hmp3}. 
The neutron EDM was estimated to be  
in the range of
$10^{-31} \sim 10^{-33}$ ecm~\cite{hmp2}. 
This is several orders of magnitude below the
experimental upper bound.  

In extensions of the SM because new sources for $CP$ violation, a non-zero
neutron
EDM can be generated at lower loop levels and therefore can be much larger than
that in the SM. 
A measurement of the neutron EDM at a level larger than $10^{-31}$ ecm would 
indicate new sources for $CP$ violation. 

\begin{figure}[htb]
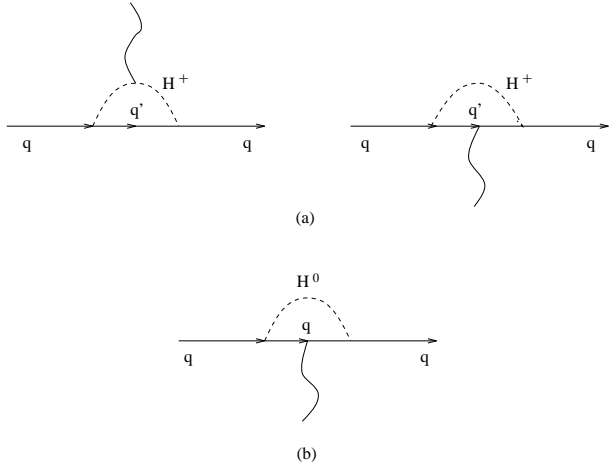

\centerline{ \DESepsf(fig-one-wein.epsf width 8 cm) }
\caption {
One loop contribution to the quark EDM in the Weinberg model.}
\label{one-w}
\end{figure}

In the Weinberg model a non-zero neutron EDM can be 
generated by exchanging Higgs particles.
At one loop level (Fig.\ref{one-w}), 
exchange of charged Higgs will generate  quark EDM's
given by~\cite{cheng1,hmp2}

\begin{eqnarray}
d_q = {G_F\over 6\sqrt{2} \pi^2} m_q Im(\alpha_1\beta_1^*)
\sum_i {x_i\over (1-x_i)^2}({3\over 4} - {5\over 4} x_i + {1-3x_i/2\over 
1-x_i} lnx_i) V_{qi}^2\;,
\end{eqnarray}
for charge -1/3 quarks, and

\begin{eqnarray}
d_q = {G_F\over 6\sqrt{2}} m_q Im(\alpha_1\beta_1^*) 
\sum_i {x_i\over (1-x_i)^2} ( x_i - {1\over 2} {1-3x_i\over 1-x_i} ln x_i) V_{iq}^2\;.
\end{eqnarray}
for charge 2/3 Quarks, where $x_i = m_i^2/m_H^2$. 
It is easy to see that $d_d >> d_u$ and the dominant contribution to $d_d$ 
is due to 
c quark in the loop. 
Taking the value of $Im(\alpha_1\beta_1)$ determined from eq.
(\ref{epsilon-w}), the neutron EDM is found to be~\cite{cheng1} 

\begin{eqnarray}
d \approx - 9 \times 10^{-26} ecm\;.
\end{eqnarray}
It is very close to the experimental upper bound. Improved measurement will
provide decisive information about this model.

The neutral Higgs contribution is predicted to be small because the quark EDM
is proportional to the third power of the 
light quark masses. However, there are
estimates which obtain larger contributions~\cite{gud}.

\begin{figure}[htb]
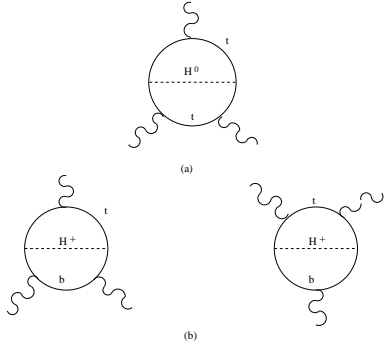

\centerline{ \DESepsf(fig-two-wein.epsf width 5 cm) }
\caption {
Two loop diagram for gluon color dipole moment.}
\label{two-w}
\end{figure}

In the Weinberg model, the contribution from
the two-loop diagrams shown in Fig.\ref{two-w}~\cite{wein-two} 
contribution to neutron EDM may dominate over that from the one loop diagrams.
The basic reason for a large EDM at two loop level is because
Higgs couplings to fermions are proportional to fermion masses. 
At one loop level,
the relevant 
fermions are the light fermions, u and d quark, but at two loop level,
heavy fermions can be in the loop, for example, the 
top quark. The couplings are much
larger which may overcome the suppression due to loop. 

One of the typical $CP$ violating operator is 

\begin{eqnarray}
O = -{1\over 6} f_{\alpha\beta\gamma}G_{\alpha\mu\rho}G_{\beta\nu}^\rho 
G_{\gamma\lambda\sigma}\epsilon^{\mu\nu\lambda\sigma}\;.
\end{eqnarray}

It is, of course, very difficult to calculate its contribution to 
the neutron EDM.
The NDA estimate gives

\begin{eqnarray}
d = {e\over 4\pi} \Lambda_\chi C\;,
\end{eqnarray}
where $C$ is the coefficient of $O$.

The neutral Higgs contribution to $C$ is give by~\cite{wein-two}
\begin{eqnarray}
C = \xi(\mu) {G_F\over 2\sqrt{2}} ImZ_1 h(m^2_t/m_H^2)\;,
\end{eqnarray}
where the function $h(x)$ is from the loop integral and is given by

\begin{eqnarray}
h(x) = {1\over 4} \left ({m_t\over m_{H_1}}\right )^4 \int^1_0 dx \int^1_0
{x^3y^3(1-x)\over ((m_t/m_{H_1})^2x(1-xy) +(1-x)(1-y))^2}\;,
\end{eqnarray}
$\xi(\mu)$ is the QCD correction factor which is given by~\cite{b-l}
\begin{eqnarray}
\xi(\mu) = \left ({g_s(m_b)\over g_s(m_t)}\right )^{-18/\beta_5}
\left ( {g_s(m_c)\over g_s(m_b)}\right )^{-18/\beta_4} \left ( 
{g_s(\mu)\over g_s(m_c)}\right )^{-18/\beta_3}\;,
\end{eqnarray}
and $ImZ_1$ is the $CP$ violating parameter in the Higgs propagator which is
proportional to $\delta_1 \rho_1$.

Assuming $m_t \sim m_{H_1}$, and taking $g_s(\mu) = 4\pi /\sqrt{6}$, one obtains

\begin{eqnarray}
d \approx 4\times 10^{-26} Im Z_1 ecm\;.
\end{eqnarray}

There is also charged Higgs contribution. 
It is given by~\cite{wein-two} 

\begin{eqnarray}
d = 4\times 10^{-24} Im(\alpha_1\beta^*_1) h_c((m_t/m_{H_1})^2)\;,
\end{eqnarray}
with
\begin{eqnarray}
h_c(x) \approx {1\over 4}{x\over (1-x)^3} (-ln x -{3\over 2} + 2 x - {1\over 2} x^2)\;.
\end{eqnarray}

If one uses the value for $Im(\alpha_1\beta_1^*)$ from eq.(\ref{epsilon-w}) 
the neutron EDM is
above the experimental upper bound. However, one must be very careful to draw
conclusion that the Weinberg model is ruled out from this consideration. 
As has been pointed out previously that
the NDA estimate is just an order of magnitude guess. A factor of ten can be 
easily missed. The above estimate may be just such a case.

\begin{figure}[htb]
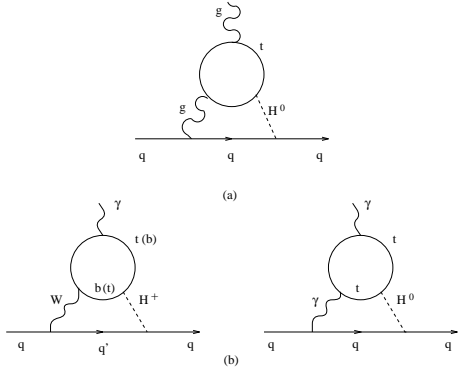

\centerline{ \DESepsf(fig-two-bz.epsf width 6 cm) }
\caption {
Additional two loop diagrams.}   
\label{two-bz}
\end{figure}

At two loop level, there are several other contributions which can generate 
a neutron EDM 
close to the experimental upper limit~\cite{bz,gunion,hmp4}. 
Some of them are shown in 
Fig.\ref{two-bz}~\cite{bz,gunion}.

\begin{figure}[htb]
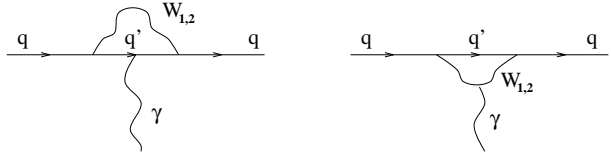

\centerline{ \DESepsf(fig-edm-lr.epsf width 8 cm) }
\caption {
One loop diagrams for quark EDM in Left-Right symmetric model.}
\label{one-lr}
\end{figure}

In the Left-Right
symmetric model due to mixing between left-handed and right-handed charged 
currents, the quark EDM is 
generated at one loop level as shown in Fig. \ref{one-lr}.
One has~\cite{hmp1,l-r-edm} 

\begin{eqnarray}
d&&\approx sin2\xi [4.5 sin(\gamma - \delta_2) + 74 sing(\gamma+\delta_1)
\nonumber\\
&& -
1.1 sin(\gamma-\delta_1) + 16 sin(\gamma+ \delta_2)] \times 10^{-23} ecm\;.
\end{eqnarray}
The neutron EDM from this contribution can be close to the experimental upper
bound.

\subsection{The Electron EDM}

The best limit on the electron EDM $d_e$ is from the EDM 
measurement of $^{205}Tl$, 
and  is given by~\cite{e-edm}

\begin{eqnarray}
d_e = [1.8\pm 1.2(stat.) \pm 1.0(sys.)]\times 10^{-27} ecm.
\end{eqnarray}

In the SM the electron EDM $d_e$ is zero at three loop level~\cite{k-p} 
and is predicted to
be less than
$10^{-38}$ ecm.

In the  Weinberg model $d_e$ is generated at one loop level. 
However at this level, $d_e$ 
is proportional to the third power in the electron mass. $d_e$ is 
predicted to be  very small ($10^{-33} ecm$). 
At two loop level, $d_e$ can be quite large due to
the second diagram (b) in Fig.\ref{two-bz} with quarks replaced by leptons 
in the diagram. 
$d_e$ can be close to the experimental bound
~\cite{bz}.

In the Left-Right symmetric model, $d_e$ is also generated at one 
loop level. It can be as large as $10^{-27}$ ecm~\cite{b-s}.

The measurements of EDM's of neutron and electron are very interesting 
measurements because in the SM the EDM's 
for these particles are predicted to be very small. 
Any new measurements with improved sensitivity may reveal 
new source for $CP$ violation. In fact present upper bound 
provided very strong constraints on models as 
discussed before. There is a potential problem for the Weinberg
model. Another interesting problem related to the neutron EDM measurement
is the so called strong $CP$ problem. This problem will be briefly discussed
in the following section.

\subsection{The strong $CP$ problem}

It has long been realised that due to instanton effects~\cite{strong-instanton}
 in non-Abelian gauge
theory, the total divergence term

\begin{eqnarray}
{1\over 2} \epsilon_{\mu\nu\alpha\beta} G^{\mu\nu}_a G_{\alpha\beta}^a = \tilde
G_a^{\alpha\beta} G^a_{\alpha\beta}\;.
\end{eqnarray}
constructed from the field 
strength $G^{\mu\nu}_a$ has non-vanishing physical effects. 
The index "a" is an internal group index.
In the case of $QCD$, $G^{\mu\nu}_a$ is the gluon field strength. The full
QCD Lagrangian is given by

\begin{eqnarray}
L_{QCD} = -{1\over 4} G^{\mu\nu}_aG^a_{\mu\nu} + \bar q (D_\mu \gamma^\mu -m) q 
-\theta {g^2\over 32\pi^2} \tilde G^{\mu\nu}_a G_{\mu\nu}^a\;,
\end{eqnarray}
where q is the quark field, m is the quark mass, $D_\mu$ is the covariant
derivative and $\theta$ is a constant. 

\begin{figure}[htb]
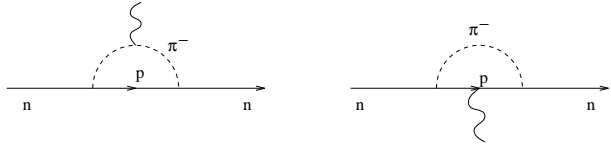

\centerline{ \DESepsf(fig-strong.epsf width 8 cm) }
\caption {
One loop diagrams for the neutron EDM due to the strong $\theta$ term.}
\label{strong}
\end{figure}

The last term in $L_{QCD}$ violates $P$ and $CP$. This term will generate $CP$
violating nucleon-meson interaction at low energy. Using chiral 
realisation of this interaction, one obtains~\cite{crewther}

\begin{eqnarray}
&&L_{\pi NN} = \sqrt{2}  \bar N 
\vec \pi \cdot \vec \tau (i\gamma_5 g_{\pi NN} + f_{\pi NN}) N\;,\nonumber\\
&&f_{\pi NN} = -{1\over \sqrt{2}} 2(m_\Xi - m_\Sigma) {m_um_dm_s \over
F_\pi (m_u+m_d)(2m_s - m_u -m_d)}\;.
\end{eqnarray}
Here $g_{\pi NN}^2/4\pi \approx 14$ is the $CP$ conserving strong nucleon-meson 
coupling constant, and 
$F_\pi = 93$ MeV is the pion decay constant.
At one loop level a non-zero neutron EDM is generated. The result obtained 
from the Feynman
diagrams in Fig.\ref{strong} is~\cite{crewther}

\begin{eqnarray}
d = -3.8 \times 10^{-16}\theta \mbox{ecm}\;.
\end{eqnarray}
And the full one loop result is given by~\cite{hmp2}

\begin{eqnarray}
2.5 \times 10^{-16}\theta \mbox{ecm} < |d|< 4.6\times 10^{-16}\theta\mbox{edm}\;.
\end{eqnarray}
The experimental upper limit on the neutron EDM implies that
$\theta$ must be less than $3\times 10^{-10}$. A coupling constant appearing 
in QCD is expected to be a much larger number. A dimensionless number as small
as $10^{-9}$ is un-naturally small. This is the strong $CP$ problem. 
Many attempts have been made to explain the smallness of the $\theta$ 
parameter or 
to
make it automatically zero. Solutions include zero u-quark
mass, axion models~\cite{pq} and etc which I will not discuss here. 

\section{Partial Rate Asymmetry}

The partial rate asymmetry is defined as 

\begin{eqnarray}
\Delta = {\bar \Gamma - \Gamma\over \bar \Gamma +\Gamma}\;,
\end{eqnarray}
where $\Gamma$ and $\bar \Gamma$ are the decay widths for a particle and its 
anti-particle, respectively. 
A non-zero $\Delta$ signals $CP$ violation. For a decay process which has two
component amplitudes, the decay amplitudes for
$A$ and $\bar A$ can be written as 

\begin{eqnarray}
A&=& A_1 e^{i(\delta^s_1+\delta_1^w)} + A_2 e^{i(\delta^s_2+\delta_2^w)}\;,
\nonumber\\
\bar A&=& A_1 e^{i(\delta^s_1-\delta_1^w)} + A_2 e^{i(\delta^s_2-\delta_2^w)}\;,
\end{eqnarray}
where $\delta_i^s$ are the strong
phases and $\delta_i^w$ are the weak phases.

Expressing the rate asymmetry $\Delta$ in terms of the quantities in the decay
amplitudes, one obtains 

\begin{eqnarray}
\Delta = {2A_1A_2 sin(\delta_1^w-\delta_2^w) sin(\delta_1^s-\delta_2^s)\over
A_1^2+A_2^2 +2 A_1A_2 cos(\delta_1^w-\delta_2^w)cos(\delta_1^w-\delta_w^2)}\;.
\end{eqnarray}

It is clear that in order to have a non-zero rate asymmetry the two
component amplitudes must have different weak and strong phases. More generally,
in order to have a non-zero rate asymmetry, 
there must exist at least
two component decay 
amplitudes with at least two different weak and strong phases.

In the SM the weak phases are due to $CP$ violation in the KM matrix. The strong
phases are difficult to calculate. This is particularly
true for exclusive decays. However, for inclusive decays the situation may
be slightly better. The strong phases obtained at quark level
may be a good representation of the size and the sign 
of the phases by appealing to
the 
quark-hadron duality. 

As an example let me consider $B^-\rightarrow K^- X$ in
the SM, where $X$ only contains $u$ and $d$ quarks~\cite{bdhp} . The effective Hamiltonian
for this decay is given in Section 3. The strong phases are generated in the
penguin diagrams when the internal quarks are $u$ and $c$.  
The typical branching ratio for this decay is about $10^{-4}$ 
and the partial rare 
asymmetry is typically -10\%~\cite{bdhp} with a cut on the kaon energy
$E_K>2$ GeV for easy experimental measurement. 
This will be measured at $B$ factories. 
There are similar calculations for exclusive decays. The asymmetries in some 
decays can be large~\cite{exclusive}. 

In the Weinberg and Left-Right symmetric models the asymmetries, 
depending on the 
detailed parameters of the model, can be the same order of magnitude as the 
SM or even larger.

\section{Test Of $CP$ Violation Involving Polarisation
Measurement}

As an example in this category of $CP$ violation measurement, which is also interesting for the Beijing 
$e^+e^-$ collider, I consider a neutral 
vector meson decay into two fermions. Let 
$\vec p_f$ and $\vec p_{\bar f}$ be the spatial momenta for the particle $f$ 
and 
its anti-particle $\bar f$ in the rest frame of the decaying vector meson $V$, 
$\vec s_f$ and $\vec s_{\bar f}$ be the polarisation of
the final fermions in their rest frames, respectively. One can construct
$CP$ violating observable from these  quantities. For example

\begin{eqnarray}
O = {1\over 2} (\hat p_f - \hat p_{\bar f})\cdot (\vec s_f \times \vec s_{\bar f})\;,
\end{eqnarray}
where $\hat p$ is the unit direction vector of the momentum.
Under $CP$ transformation, $O\rightarrow -O$. If the average value
$<O>$ is not zero, it signals $CP$ violation.

\subsection{$CP$ violation in $J/\psi \rightarrow \Lambda \bar \Lambda$}

The decay amplitude for this process can be parameterized as 
\begin{eqnarray}
A(J/\psi\rightarrow \Lambda \bar \Lambda) = \epsilon_\mu \bar u(p_\Lambda)
[\gamma_\mu(a+ b\gamma_5) + (p_\Lambda - p_{\bar \Lambda})_\mu (c+id\gamma_5)]
u(p_{\bar \Lambda})\;,
\end{eqnarray}
where $\epsilon^\mu$ is the polarisation vector of $J/\psi$. A 
non-zero value for
$d$ violates $CP$ in this decay. One finds~\cite{hmm}

\begin{eqnarray}
<O> = {(1-4m^2_\Lambda / m^2_\psi) m_\psi(2m_\psi a +(m^2_\psi - 4 m^2_\Lambda)
c)\over 54\pi \Gamma(J/\psi \rightarrow \Lambda \bar \Lambda)} d\;.
\end{eqnarray}

The main decay channel, $\Lambda (\vec s_1) \rightarrow p(\vec q_1)
+\pi^-$ and its 
anti-particle decay can be used to measure the polarisation for $\Lambda$ and 
$\bar \Lambda$. The density matrices for these decays in the
rest frame of $\Lambda$ and $\bar \Lambda$ are of the form 

\begin{eqnarray}
\rho_\Lambda &=& 1+\alpha_- \vec s_1 \cdot \hat q_1\;\;\;\mbox{for $\Lambda$ decay}\;,
\nonumber\\
\rho_{\bar \Lambda} &=& 1+\alpha_+ \vec s_2 \cdot \hat q_2\;\;\;\mbox{for $\bar 
\Lambda$ decay}\;,
\end{eqnarray}
where $\alpha_-\approx \alpha_+ = 0.642\pm 0.013$. 
A more convenient experimental measurable can be
defined, $\tilde O = \hat p_\Lambda \cdot (\hat q_1\times q_2)$. The relation 
between $<\tilde O>$ and $<O>$ is given by 

\begin{eqnarray}
<\tilde O> = \alpha_-\alpha_+ <O>\;.
\end{eqnarray}

To have an idea how sensitive this measurement can give information for 
fundamental quantities, I consider the case that the parameter $d$ is due to the
$\Lambda$ EDM, $d_\Lambda$. 
Exchanging a photon between a charm quark and a $\Lambda$, one
obtains
\begin{eqnarray}
d = -{2\over 3} {g_V\over m^2_\psi} e d_\Lambda\;,
\end{eqnarray}
where the parameter $g_V$ is defined as $<0|\bar c\gamma_\mu c|J/\psi> = \epsilon_\mu g_V$, which is determined to be $1.25$ GeV$^2$ from $J/\psi \rightarrow
\mu^+\mu^-$. 

The decay is expected to be dominated by $CP$ and $P$ conserving
amplitudes, 
$a$ and $c$. Due to large errors associated with the data it is not possible
to separately determine $a$ and $c$. Two representative cases will be 
considered: 1) $a$ term 
dominates the decay; and b) $b$ term dominates the decay.
The results are~\cite{hmm}

\begin{eqnarray}
<\tilde O> = \left \{ \begin{array}{ll}
3.17\times 10^{-3} d_\Lambda /(10^{-16} \mbox{edm}),&\mbox{if a term 
dominates;}\\
0.71\times 10^{-3} d_\Lambda /(10^{-16}\mbox{ecm}),& \mbox{if b
 term dominates.}
\end{array}
\right .
\end{eqnarray}

The present upper bound on $d_\Lambda$ is $1.5\times 10^{-16} ecm$. It is 
possible to improve this bound with more than $10^{6}$ $\Lambda,\bar 
\Lambda$ pairs which will require
at least $10^9$ $J/\psi$'s. This may only be achieved at a $J/\psi$ factory. 
However, similar analysis can be carried out for 
other hyperons, 
such as $\Sigma$ and $\Xi$. It is
possible to obtain interesting bounds on the EDM's of these hyperons even 
using the available data at the Beijing $e^+e^-$ collider. 

The EDM's for  $\Lambda$, $\Sigma$ and $\Xi$ are all much smaller than 
$10^{-16}$ ecm in the SM, the Weinberg and the Left-Right symmetric models.
Measurements of $<\tilde O>$ at a level of $10^{-3}$ will certainly indicate
new physics beyond the SM and beyond the models discussed here.

\subsection{$CP$ violation in $e^+e^-\rightarrow \tau^+\tau^-$}

Another interesting experiment for $CP$ violation 
may be performed at the Beijing $e^+e^-$ collider is the measurement of a
$CP$ violating observable $<T>$ related to $\tau$ production and its decays,
which is defined by~\cite{b-n-o}

\begin{eqnarray}
T^{ij} = (\hat q_+ - \hat q_-)^i {(\hat q_+ \times \hat q_-)^j\over 
|\hat q_+ \times \hat q_-|}\;,
\end{eqnarray}
where $\hat q_+$ and $\hat q_-$ are the directions of the momenta of the
final states from $\tau^+ \rightarrow A(q_+) + X_+$ and $\tau^-\rightarrow
B(q_-) +X_-$, respectively. Here the tauons are produced in the process
$e^+e^- \rightarrow \tau^+\tau^-$. If the average value of $T$, $<T>$ is
non-zero, $CP$ is violated. 

Assuming $CP$ violation is purely from the tauon EDM $d_\tau$, $<T>$ is 
given by~\cite{huang}

\begin{eqnarray}
<T^{ij}> &=& {E_{cm}\over e} d_\tau C_{AB} diag(-1/6, -1/6, 2/6)\nonumber\\
&=& 1.7\times 10^{-3} {d_\tau\over 5\times 10^{-17}(ecm)} {E_{cm}\over 4 (GeV)}
C_{AB} (-1,-1,2)\;,
\end{eqnarray}
where $E_{cm}$ is the energy in the $e^+e^-$ central mass frame, and 
$C_{AB}$ is of order one depending on the specific final states
$A$ and $B$.

At present $d_\tau$ is bounded to be less than~\cite{pdg}
 $5\times 10^{-17}$ecm. With
$10^6$ $e^+e^-\rightarrow \tau^+\tau^-$ events, it is possible to improve the
limit on $d_\tau$. This may be achieved at tau-charm factories. 

Theoretical predictions for $d_\tau$ is extremely small in the 
SM~\cite{huang}. In 
the Weinberg and Left-Right symmetric models, it can be as large as $10^{-18}$
ecm. Any measurement of $<T^{ij}>$ at order $10^{-3}$ will, again, 
indicate new physics
beyond the SM and the models discussed here.

\subsection{$CP$ violation in hyperon decays}

As a final example of studying 
$CP$ violation involving polarisation measurement, 
I discuss 
the E871 experiment at Fermilab~\cite{e871}
which measures polarisations in $\Xi\rightarrow 
\Lambda \pi \rightarrow p \pi \pi$
decay.
The quantity to be measured is

\begin{eqnarray}
A = A_\Lambda + A_\Xi\;,\;\;
A_\Lambda = {\alpha_\Lambda + \alpha_{\bar \Lambda}\over \alpha_\Lambda - 
\alpha_{\bar \Lambda}}\;,\;\; A_\Xi = {\alpha_\Xi +\alpha_{\bar \Xi}\over
\alpha_\Xi - \alpha_{\bar \Xi}}\;,
\end{eqnarray}
where $\alpha_i$ are the polarisation constants in $i \rightarrow j \pi$ decays.
A non-zero value for $A_i$ implies $CP$ violation.
The sensitivity of E871 will reach $10^{-4}$. In the SM this quantity is
predicted to be a few times $10^{-5}$ with a smaller number in the Weinberg 
model~\cite{dhp,hv}. In Left-Right symmetric model, 
this quantity can be as large as~\cite{chp} a few times 
$10^{-4}$
which will be probed by the E871 experiment. Useful information
about $CP$ violation will be obtained in this experiment.

\section{Baryon Number Asymmetry}

An important fact of the universe is 
that our local region consists primarily of matter and not anti-matter.
There is an asymmetry in baryon number.  
The baryon number 
asymmetry appears nearly maximal, that is, there are hardly any anti-baryons. 
However, when viewed from the perspective of cosmology it is actually 
very tiny. Analysis of nucleosynthesis of light element of the universe
gives~\cite{pdg}
\begin{eqnarray}
{n_B -n_{\bar B}\over n_\gamma} \sim 10^{-10} - 10^{-9}\;,
\end{eqnarray}
about the same as or more than the net baryon density inferred from the visible
matter of the universe.
Here $n_{B,\bar B, \gamma}$ are respectively the averaged number densities of 
baryons, anti-baryons and microwave photons in the present universe. If the
baryon number were exactly conserved and if the initial net baryon number of the
universe were zero, one would expect $n_B = n_{\bar B}$, and 
baryon and anti-baryon present in the early 
universe would have been almost 
annihilated, producing a very small residual matter and
anti-matter, 
almost nine orders of magnitude too small
$(n_B, n_{\bar B})/n_\gamma \le 10^{-19}$. Therefore the 
observed baryon number asymmetry 
would probably have to be postulated as an
initial condition on the big bang if the baryon number were conserved.
Such a small number as an initial condition is 
possible but very unesthetic. 
More elegant scenario is possible.
It was shown by Sakharov in 1966~\cite{sakh} that it is possible 
for the universe to have 
an initially zero net  baryon number to evolve to our present universe with 
baryon number asymmetry generated dynamically after big bang 
if the following three
conditions are satisfied:
\begin{itemize}

\item Baryon number violating 
interaction; 
\item $C$ and $CP$ violating interaction;
\item  
Deviation from thermal equilibrium.
\end{itemize}
These conditions are all crucial for the generation of baryon number asymmetry
in the universe.
(i) If baryon number were conserved, 
the universe would be symmetric, rather than asymmetric in baryon
number. 
(ii) 
If $C$ or $CP$ were conserved, then the rate of reactions with particle would 
be the same as that of its anti-particle. 
No charge asymmetry could develop from it.
(iii)
If the universe is always in 
thermal equilibrium with zero initial baryon number, 
then it is zero forever.
It is clear that baryon number 
asymmetry 
signals $CP$ violation.

Many theoretical efforts have been made to build concrete models to realise 
the necessary conditions and produce required baryon number 
asymmetry. 
The SM has all the ingredients to generate baryon number asymmetry in the 
universe with baryon number violation from the anomalous 
$SU(2)_L$ interaction, 
$CP$ violation 
from the $KM$ matrix and deviation from thermal equilibrium from the symmetry 
breaking
phase transition~\cite{rev}. However, 
it is believed that 
the SM alone
does not provide enough $CP$ violation to explain the baryon number asymmetry
~\cite{rev}. 
One has to go beyond the SM.
Other mechanisms for $CP$ violation may be in 
operation, such as $CP$ in the Higgs interaction in the Weinberg model.
This provides another reason for study $CP$ violation beyond the SM. The 
study of baryon number asymmetry will help to understand the origin of
$CP$ violation.

\section{Conclusion}

More than 30 years have passed since the surprising discovery of $CP$ violation
in neutral kaon system in 1964, the origin of $CP$ violation is still a mystery.
Many models have been proposed to explain the observed $CP$ violation in
$K^0-\bar K^0$ mixing. 
>From previous discussions, it is clear that the SM is consistent with all
laboratory experimental data. However there are also extensions of the SM
which can equally well explain experimental data. No satisfactory explanation
for $CP$ violation has been established. More experiments are needed to pin down
the origin of $CP$ violation. Many experiments have been  and are being carried 
out.
Although no new signal for $CP$ violation has 
been observed in laboratory systems, 
considerable progress have been made in obtaining limits on various
experimental measurables, for example, $\epsilon'/\epsilon$, the EDM's of
neutron and electron, and etc. These bounds have put interesting constraints
on theoretical models. New experiments at
$B$ factories, and other facilities, will provide more and decisive information
about $CP$ violation. It is hopeful that the origin of $CP$ violation will 
finally be understood.

\section*{Acknowledgments} 
This work was supported in part by the Australian Research Council, by the 
KC. Wong Education Foundation, Hong Kong, and by the Australian Academy of 
Science. I thank T. Browder, A. Datta, N. Deshpande, J. Donoghue, 
J. Ma, B. McKellar, S. Oh, S. Pakvasa, H. Steger and G. Valencia for 
    collaborations in 
related topics.  I also thank Dr. J.P. Ma for many useful discussions,
 and 
the CCAST and the 
Institute of Theoretical Physics, China for hospitality where 
part of this work was done.


\begin{thebibliography}{99}

\bibitem{noether} E. Noether, Nachr. Klg. Ges. Wiss. G\"ott. Math. Physik, Kl. 235(1918).

\bibitem{stu-feyn} E.C.C. St\"uckelberg, Helv. Phys. Acta, {\bf 12}, 23(1942); 
R. Feynman, Phys. Rev. {\bf 74}, 939(1948); {\bf 76}, 749(1949).

\bibitem{lee-yang} T.D. Lee and C.N. Yang, Phys. Rev. {\bf 104}, 254(1956).

\bibitem{wu} C.S. Wu, R.W. Hayward, D.D. Hoppes and R.P. Hudson, Phys. Rev. 
{\bf 105}, 1413(1957).

\bibitem{parity} R.L. Garwin, L.M. Lederman and M. Weinrich, Phys. Rev. {\bf 105}, 
1415(1957); J.J. Friedman and V.L. Telegdi, Phys. Rev. {\bf 105}, 1681(1957). 

\bibitem{cp-exp} J.H. Christenson, J.W. Cronin, V.L. Fitch and R. Turlay, 
Phys. Rev. Lett. {\bf 13}, 138(1964).

\bibitem{cp-rev} C. Jarlskog, CP Violation, World Scientific, Singapore (1989);
L. Wolfenstein, CP Violation, North-Holland, Amsterdam, Netherland (1989).

\bibitem{wigner} E.P. Wigner, G\"ott. Nach. Math. Naturw. Kl., P 546(1932). 


\bibitem{cpt} J. Schwinger, Phys. Rev. {\bf 82}, 914(1951); {\bf 91}, 713(1953);
G. L\"uders, Kgl. Danske Videnskab Selskab. Mat. Fys. Medd. 28, No. 5 (1954);
W. Pauli, In Niels Bohr and the Development of Physics, Ed. W. Pauli, 
McGraw-Hill, New York, 1955; R.F. Streater and A.S. Wightman, PCT, Spin and 
Statistics, and All That,  Benjamin, New York, 1964.

\bibitem{t-cpt} T.D. Lee, R. Oehme 
and C.N. Yang, Phys. Rev. {\bf 106}, 340(1957);
T.D. Lee and C.S. Wu, Annu. Rev. Nucl. Sci. {\bf V16}, 471(1966).

\bibitem{pdg} Particle Data Group (R. Barnett et al.), 
Phys. Rev. {\bf D54}, 1(1996). 

\bibitem{super} L. Wolfenstein, Phys. Rev. Lett. {\bf 13}, 562(1964);
T.D. Lee and L. Wolfenstein, Phys. Rev. {\bf B138}, 1490(1965). 

\bibitem{spon} T.D. Lee, Phys. Rev. {\bf D8}, 1226(1973); 
Phys. Rep. {\bf 96}, 143(1976).

\bibitem{weinberg} S. Weinberg, Phys. Rev. Lett. {\bf 31}, 657(1976).

\bibitem{l-r} R.N. Mohapatra and J.C. Pati, Phys. Rev. {\bf D11}, 566(1975).

\bibitem{km} M. Kobayashi and K. Maskawa, 
Progr. Theor. Phys. {\bf 49}, 652(1973). 

\bibitem{w-sm} S.L. Glashow, Nucl. Phys. {\bf 22}, 579(1961);
S. Weinberg,Phys. Rev. Lett. {\bf 19}, 1264(1967);
A. Salam, Proceedings of Eighth Nobel Symposium, Ed. N. Svartholm,
Wiley-Interscience, New York, 1968.

\bibitem{s-sm} H.D. Politzer, Phys. Rev. Lett. {\bf 30}, 1346(1973);
D. Gross and F. Wilczek, Phys. Rev. Lett. {\bf 30}, 1343(1973).

\bibitem{lep} S.C. Ting, In proceedings of the 17th Int. Sym. on Lepton-
Photon interactions, Aug. 10-15, 1995, Beijing, China. World Scientific, 
Singapore (1996).

\bibitem{nucl-syn} K. Olive and D. Thomas, Astropart. Phys. {\bf 7}, 27(1997).

\bibitem{breaking} For a recent review see: R.S. Chivukula, preprint, hep-ph/9701322.

\bibitem{higgs} P. Higgs, Phys. Lett. {\bf 12}, 132, {\bf 13}, 508(1964);
Phys. Rev. {\bf 145}, 1156(1966);
G.S. Guralnik, C.R. Hagen and T.W. Kibbile, Phys. Rev. Lett. {\bf 13}, 585(1964);
F. Englert and R. Brout, Phys. Rev. Lett. {\bf 13}, 321(1964).

\bibitem{higgs-mass} J. Gunion, A. Stange, S. Willenbrock, preprint, hep-ph/9602238. 

\bibitem{wolf} L. Wolfenstein, Phys. Rev. Lett. {\bf 51}, 1945(1983).

\bibitem{vud} I.S. Towner, Nucl. Phys. {\bf A540}, 478(1992).

\bibitem{vus} H. Leutwyler and M. Roos, Z. Phys. {\bf C25}, 91(1984);
J. Donoghue, B. Holstein and S. Klimt, Phys. Rev. {\bf D35}, 934(1987).


\bibitem{vcb} L. Gibbons, In proceedings of the XXVIII int. conf. on high
energy physics, Warsaw, July 25-31, 1996.

\bibitem{vub} M. Neubert, Int. J. Mod. Phys. {\bf A11}, 4173(1996).

\bibitem{box} T. Inami and C.S. Lim, Prog. Theor. Phys. {\bf 65}, 297(1981);
ibid, 1772(E).

\bibitem{rosner} J. Rosner, preprint, hep-ph/9612327; A. Ali and D. London,
preprin, hep-ph/9607392.

\bibitem{qcd} A.J. Buras, M. Jamin and P. Weisz, Nucl. Phys. {\bf B347}, 
491(1990); S. Herrlich and U. Nierste, Nucl. Phys. {\bf B419}, 292(1994);
Phys. Rev. {\bf D52}, 6505(1995).

\bibitem{top} F. Abe et al., Phys. Rev. Lett. {\bf 74}, 2626(1995); 
S. Abchi et al., ibid. 2632(1995).

\bibitem{k-l}  R. Aleksan, B. Kayser and D. London, Phys. Rev. Lett. {\bf 73},
18(1994).

\bibitem{methods} Y. Nir and H. Quinn, in $B$ Decay, ed. S. Stone, P.520,
World Scientific, Singapore (1994); I. Dunietz, ibid. p550;
M. Gronau, D. London, Phys. Rev. Lett. {\bf 73}, 21(1994); M. Gronau,
O. Hernandez, D. London and J. Rosner, Phys. Rev. {\bf D50}, 4529(1994);
Phys. Lett. {\bf B333}, 500(1994); {\bf D52}, 6411(1995); 
J. Silva and L. Wolfenstein, Phys. Rev. {\bf D49}, R1151(1994). 
C. Hamzaoui and Z.-Z. Xing, Phys. Lett. 
{\bf B360}, 131(1995);
A. Buras and R. Fleischer, Phys. Lett. {\bf B360}, 138(1995).
 N.G. Deshpande and Xiao-Gang He, Phys. Rev. Lett. {\bf 75}, 3064(1995);
A.S. Dighe, M. Gronau and J. Rosner, Phys. Rev. {\bf D54}, 4677(1996);
N. Deshpande, Xiao-Gang He and Sechul Oh, Z. Phys. {\bf C74}, 359(1997).

\bibitem{h-b} A. Buras, M. Jamin, M. Lautenbacher and P. Weisz, Nucl. Phys.
{\bf B400}, 37(1993); A. Buras, M. Jamin and Lautenbacher, ibid, 75(1993);
M. Chiuchini, E. Franco, G. Marinelli and L. Reina, Nucl. Phys. {\bf B415}, 
403(1994).

\bibitem{d-h1} 
R. Fleischer, Z. Phys. {\bf C58}, 438(1993); ibid. {\bf C62}, 81(1994);
G. Kramer, W. Palmer and H. Simma, Nucl. Phys. {\bf 428}, 77(1994);
Z. Phys. {\bf C66}, 429(1994);
N. Deshpande and Xiao-Gang He, Phys. Lett. {\bf B336}, 471(1994).

\bibitem{b-factory} Letter 
of intent for the study of CP violaiton and heavy flavor
physics at PEPII, The BABAR Collaboration, 1994; Letter of intent 
for a study of CP
violation in B meson decays, KEK Report 94-2, April 1994.

\bibitem{stone} $B$ Decays, Ed. S. Stone, World Scientific, Singapore, 1994.



\bibitem{sanda} A.B. Carter and A. I. Sanda, Phys. Rev. Lett. {\bf 45}, 952(1980);
Phys. Rev. {\bf D23}, 1567(1981); I.I. Bigi and A.I. Sanda, 
Nucl. Phys. {\bf B193}, 85(1981); ibid, {\bf B281}, 41(1987).

\bibitem{gl} M. Gronau and D. London, Phy. Rev. Lett. {\bf 65}, 3381(1990).

\bibitem{d-h2} N. Deshpande and Xiao-Gang He, Phys. Rev. Lett. {\bf 74}, 26(1995);
ibid 4099(E).

\bibitem{desh-he} N. Deshpande and Xiao-Gang He, Phys. Rev. Lett. {\bf 75}, 1703(1995).

\bibitem{quin} A. Snyder and H. Quinn, Phys. Rev. {\bf D48}, 2139(1993).

\bibitem{fles} R. Fleischer, preprint, hep-ph/9612446.

\bibitem{wyler} M. Gronau and D. Wyler, Phys. lett. {\bf B256}, 172(1991).

\bibitem{soni} I. Dunietz, Phys. Lett. {\bf B270}, 75(1991);
R. Aleksan, I. Dunietz and B. Kayser, Z. Phys. {\bf C54}, 653(1992);
D. Atwood, I. Dunietz and A. Soni, Phys. Rev. Lett. {\bf 78}, 3257(1997).



\bibitem{d-b} N. Deshpande and B. Dutta, Phys. Rev. Lett. {\bf 77}, 4499(1996);
A. Cohen, D. Kaplan, F. Lepeintre and A. Nelson, Phys. Rev. Lett. {\bf 78},
2300(1997).

\bibitem{cheng1} H.-Y. Cheng, Int. J. Mod. Phys. {\bf A7}, 1059(1992).

\bibitem{cp-two} G. Branco, A. Buras and J.-M. Gerard, Nucl. Phys. {\bf B259}, 
306(1985).

\bibitem{wolf-liu} J. Liu and L. Wolfenstein, Nucl. Phys. {\bf B289}, 1(1987).

\bibitem{hall-w} L. Hall and S. Weinberg, Phys. Rev. {\bf D48}, 979(1993).

\bibitem{l-wu} Y.-L. Wu and L. Wolfenstein, Phys. Rev. Lett. {\bf 73}, 1762(1994).


\bibitem{branco} G. Branco, Phys. Rev. Lett. {\bf 44}, 504(198); Phys. Rev.
{\bf D22}, 2901(1980). 

\bibitem{tye} C. Albright, J. Smith and S.-H. Tye, Phys. Rev. {\bf D21}, 711
(1980); K. Shizuya and S.-H. Tye, Phys. Rev. {\bf D23}, 1613(1981).

\bibitem{s-d} A.I. Sanda, Phys. Rev. {\bf D23}, 2647(1981); N. Deshpande, ibid.
2654.

\bibitem{dhg} J. Donoghue and B. Holstein, Phys. Rev. {\bf D32}, 1152(1985). 

\bibitem{cheng} H.-Y. Cheng, Phys. Rev. {\bf D34}, 1397(1986).

\bibitem{chau-cheng} J. Donoghue, B. Holstein and Y.C. Lin, Nucl. Phys.
{\bf B277}, 651(1986);
H.Y. Cheng, Phys. Lett. {\bf B245}, 122(1990).

\bibitem{ll-r} R. Mohapatra and G. Senjanovic, Phys. Rev. Lett. {\bf 40},
912(1980); Phys. Rev. {\bf D23}, 165(1981).

\bibitem{soni-l} G. Beal, M. Bander and A. Soni, Phys. Rev. Lett. {\bf 48}, 848(1982).

\bibitem{chang} D. Chang, J. Basecq, L.F. Li and A. Soni, Phys. Rev.{\bf D30},
1601(1984); W.-S. Hou and A. Soni, Phys. Rev. {\bf D32}, 163(1985). 


\bibitem{hmp1} Xiao-Gang He, B. McKellar and S. Pakvasa, Phys. Rev. Lett.
{\bf 61}, 1267(1988).

\bibitem{beyond} Y. Grossman and M. Wprah, Phys. Lett. {\bf B395}, 241(1997);
Y. Grossman and Y. Nir, Phys. Lett. {\bf B407}, 307(1997).
 

\bibitem{he-w} Xiao-Gang He, Phys. Rev. {\bf D53}, 6326(1996).

\bibitem{dhg1} J. Donoghue and E. Golowich, Phys. Rev. {\bf D37}, 2542(1988). 

\bibitem{w-yang} T.T. Wu and C.N. Yang, Phys. Rev. Lett. {\bf 13}, 380(1964).

\bibitem{na31} NA31 Collaboration (G.Barr et al.), Phys. Lett. {\bf B317}, 233(1993). 

\bibitem{e731} E731 Collaboration (A. Barker et al.), Phys. Rev. Lett. {\bf 70}, 1203(1993). 

\bibitem{dhg2} J. Donoghue, E. Golowich, B. Holstein and J. Trampetic, Phys.
Lett. {\bf B179}, 361(1986); ibid. {\bf B188}, 511(1987).

\bibitem{randel} J. Flynn and L. Randall, Phys. Lett. {\bf B326}, 31(1989);
ibid. {\bf B334}, 580(E)(1990).

\bibitem{l-wu1} E. Paschos and Y.-L. Wu, Mod. Phys. Lett. {\bf A6}, 93(1991);
G. Buchalla, A. Buras and M. Harlander, Nucl. Phys. {\bf B337}, 313(1990).

\bibitem{buras-e} A. Buras, M. Jamin and M. Lantenbacher, Phys. Lett. {\bf B389}, 749(1996).

\bibitem{dane} DA$\phi$NE Handbook, Eds. L. Laiani, G. Pancheri and N. Paver
(LNF, 1994).

\bibitem{kaon-cp} For a review see J. Donoghue, B. Holstein and G. Valencia, 
Int. J. Mod. Phys. {\bf A2}, 318(1987); 
B. Winstein and L. Wolfenstein, Rev. of Mod. 
Phys. {\bf 65}, 1113(1993); J. Ritchie and s. Wojciki, Rev. of Mod. Phys. 
{\bf 65}, 1149(1993).

\bibitem{p-r} E. Pucell and N. Ramsey, Phys. Rev. {\bf 78}, 807(1950). 

\bibitem{d-exp} K. Smith et al., Phys. Lett. {\bf B234}, 234(1990);

\bibitem{e-edm} E. Commins et al., Phys. Rev. {\bf A50}, 2960(1994);
J.P. Jacobs et al., Phys. Rev. {\bf A52}, 3521(1995).

\bibitem{atom-exp} T. Vold et al., Phys. Rev. Lett. {\bf 52}, 2229(1984);
S. Murthy et al., Phys. Rev. Lett. {\bf 63}, 965(1989);
P. Cho, K. Sangster and E. Hinds, Phys. Rev. Lett. {\bf 63}, 2559(1989);
J. Jacobs et al., Phys. Rev. Lett. {\bf 71}, 3782(1993); Phys. Rev. {\bf A52}, 3521(1995).

\bibitem{hmp2} Xiao-Gang He, Bruce McKellar and Sandip Pakvasa, 
Int. J. Mod. Phys.  {\bf A4}, 5011(1989).

\bibitem{b-m} S. Barr and W. Marciano, in CP Violation, ed. C. Jarlskog,
p46, World Scientific, Singapore (1989).

\bibitem{b-s} W. Bernreuther and M. Suzuki, Rev. Mod. Phys. {\bf 63}, 313(1991);
ibid. {\bf 64}. 633(E)(1992).

\bibitem{other-edm}
V. Khatsymovsky, I. Kriplovich and A.S. Yelkhovsky, Ann. Phys. {\bf 186}, 1(1986); 
Xiao-Gang He and McKellar, Phys. Rev. {\bf D46}, 2131(1992);
Xiao-Gang He, B. McKellar and S. Pakvasa, Phys. Lett. {\bf B283}, 348(1993);
S. Barr, Int. J. Mod. Phys. {\bf A8}, 209(1993); Phys. Rev. {\bf D45}, 4148(1992);
Xiao-Gang He and B. McKellar, Phys. Rev. {\bf D47}, 4055(1993); Phys. Lett. 
{\bf B390}, 318(1997).


\bibitem{nda} A. Manohar and H. Georgi, Nucl. Phys. {\bf B234}, 189(1984).

\bibitem{sha} E.Shabalin, Sov. Phys. Usp. {\bf 26},4(1983).

\bibitem{hmp3} B. McKellar, S. Choudhury, 
Xiao-Gang He and S. Pakvasa, Phys. Lett. {\bf B197},556(1987).


\bibitem{gud} A.A. Anselm, V. Bunakov, V.P. Gudkov and N. Uraltsev, Phys. 
Lett. {\bf B152}, 116(1985).

\bibitem{wein-two} S. Weinberg, Phys. Rev. Lett. {\bf 63}, 2333(1989);
D. Dicus, Phys. Rev. {\bf D41 }, 999(1990).
 

\bibitem{b-l} E. Braaten, C.-S. Li and T. C. Yuan, Phys. Rev. Lett. 
{\bf 64}, 1709(1990).


\bibitem{bz} S. Barr and A. Zee, Phys. Rev. Lett. {\bf 65}, 21(1990).

\bibitem{gunion} J. Gunion and D. Wyler, Phys. Lett. {\bf 248}, 170(1990);
D. Chang, W.Y. Keung and T.C. Yuan, Phys. Lett. {\bf 251}, 608(1990).

\bibitem{hmp4} Xiao-Gang He, B. McKellar and S. Pakvasa, Phys. {\bf B 254}, 
231(1991).


\bibitem{l-r-edm} G. Beal and A. Soni, Phys. Rev. Lett. {\bf 47}, 552(1981).

\bibitem{k-p} M. Popelov and I. Kriplovich, Sov. J. Nucl. Phys. {\bf 53}, 
638(1991).


\bibitem{strong-instanton} A.A. Belavin, A.M. Polyakov, 
A.S. Shwarz and Y.S. Tyupkin, Phys. Lett. {\bf 59}, 85(1975);
G. 't Hooft, Phys. Rev. Lett. {\bf 37}, 8(1976); R. Jackiw and C. Rebbi, Phys.
Rev. Lett. {\bf 37}, 172(1976); C.G. Callan, R.F. Dashen and D. Gross, Phys. 
Lett. {\bf B63}, 334(1976).

\bibitem{crewther} R.J. Crewther, P. DiVecchia, G. Veneziano and E. Witten, Phys.
Lett. {\bf B88}, 123(1979).

\bibitem{pq} R. Peccei and H. Quinn, Phys. Rev. Lett. {\bf 38}, 1440(1977); Phys. Rev. {\bf D16}, 1791(1977).

\bibitem{bdhp} T. Browder, A. Datta, Xiao-Gang He and S. Pakvasa,
Preprint, hep-ph/9705320.

\bibitem{exclusive}
D.-S. Du, M.-Z. Yand and D.-Z. Zhang, Phys. Rev. {\bf D53}, 249(1996);
D.-S. Du and Li-Bo Guo, Z. Phys. {\bf C75}, 9(1997).

\bibitem{hmm} Xiao-Gang He, J.P. Ma and B. McKellar, Phys. Rev. {\bf D47},
R1744(1993); ibid. {\bf D49}, 4548(1994).

\bibitem{b-n-o} W. Bernreuther, O. Nachtmann, Phys. Rev. Lett. {\bf 63}, 
2787(1989); W. Bernreuther, O. Nachtmann, P. Overmann and T. Schroder, 
Nucl. Phys. {\bf B388}, 53(1992).

\bibitem{huang} T. Huang, W. Lu, Z-J. Tao, Phys. Rev. {\bf D55}, 1643(1997).

\bibitem{e871} G. Gidal, P.H. Ho, K.B. Luk and E.C. Dukes, Fermilab E871.

\bibitem{dhp} J. Donoghue and S. Pakvasa, Phys. Rev. Lett. {\bf 55}, 162(1985);
J. Donoghue, Xiao-Gang He and S. Pakvasa, Phys. Rev. {\bf D34}, 833(1986);
Xiao-Gang He, H. Steger and G. Valencia, Phys. Lett. {\bf B272}, 411(1991).

\bibitem{hv} Xiao-Gang He and G. Valencia, Phys. Rev. {\bf D52}, 5257(1995).

\bibitem{chp} D. Chang, Xiao-Gang He and S. Pakvasa, Phys. Rev. Lett. 
{\bf 74}, 3927(1995).

\bibitem{sakh} A.D. Sakharov, JETP Lett. {\bf 5}, 24(1967). 

\bibitem{rev} A. Cohen, D. Kaplan and A. Nelson, Ann. Rev. Nucl. Part. Sci.
{\bf V43}, 27(1993); 
V.A. Rubakov and M.E. Shaposhnikov, Phys. Usp. {\bf 39}, 461(1996).



\end{thebibliography}
\end{document}